\documentclass[10pt, conference, letterpaper]{IEEEtran}
\IEEEoverridecommandlockouts

\usepackage{float}
\usepackage{algorithmic}
\usepackage{subfigure}
\usepackage{array}
\usepackage{color}
\usepackage{graphicx}
\usepackage{balance} 
\usepackage{amssymb}
\usepackage{amsmath}
\usepackage{url}

\begin{document}
\title{No Perfect Outdoors: Towards A Deep Profiling of GNSS-based Location Contexts}

\author{Feng Li,~
        Jin Wang,~
        Jun Luo
\thanks{F. Li is with School of Computer Science and Technology, Shandong University, China. Email: fli@sdu.edu.cn.}
\thanks{J. Wang and J. Luo are with School of Computer Science and Engineering, Nanyang Technological University, Singapore. E-mail: \{jwang033, junluo\}@ntu.edu.sg}}

\maketitle

\begin{abstract}
  While both outdoor and indoor localization methods are flourishing, how to properly marry them to offer pervasive localizability in urban areas remains open. Recently proposals on indoor-outdoor detection make the first step towards such an integration, yet complicated urban environments render such a binary classification incompetent. In this paper, we intend to fully explore raw GNSS measurements in order to better characterize the diversified urban environments. Essentially, we tackle the challenges introduced by the complex GNSS data and apply a deep learning model to identify representations for respective location contexts. We further develop two preliminary applications of our deep profiling. On one hand, we offer a more fine-grained semantic classification than binary indoor-outdoor detection. On the other hand, we derive a GPS error indicator more meaningful than that provided by Google Maps. These results are all corroborated by our extensive data collection and trace-driven evaluations.
\end{abstract}


\section{Introduction} \label{sec:intro}
  Existing localization/navigation systems are clearly separated into two categories: outdoor~\cite{GPS,Hassanieh-MobiCom2012} versus indoor~\cite{Bahl-INFOCOM00,Youssef-MobiSys2005}. Normally, outdoor systems are mostly relying on GPS, while indoor systems leverage other pervasively available signal sources, such as WiFi~\cite{Bahl-INFOCOM00,Youssef-MobiSys2005,Xiong-NSDI2013}, visible light~\cite{Kuo-MobiCom2014,Zhang-MobiCom2017}, and geomagnetism~\cite{Wang-MobiSys2012,Luo-TMC15}. A strong implication behind such a clear-cut separation is the ability of differentiating indoor from outdoor, so as to trigger timely switches between these two systems; this has provoked several research proposals on detecting the transition from outdoor to indoor or vice versa (a.k.a. \textit{IO-detection})~\cite{Zhou-SenSys12,Radu-SenSys14,Chen-INFOCOM2017}. Nonetheless, though perfect indoor scenarios (wrapped by concrete walls and roof entirely) do exist, GPS signal can grow very weak while walking on downtown areas, while users inside a room with big windows may still obtain very good GPS location indicators. In short, our daily experience tells us that \textit{there is no perfect outdoor scenarios for urban areas}.

  Exactly due to these abnormal behaviors of GPS, recently researches suggest to either leverage GPS readings for indoor localization~\cite{Chintalapudi-MobiCom2010,Nirjon-MobiSys2014} or complement GPS with WiFi readings for outdoor localization~\cite{Cheng-MobiSys2005,WOLoc-INFOCOM2017}. Nonetheless, the question about how much trust we may put on each signal source for localization (in particular, to use GPS or not) remains open. Apparently, the outcome of IO-detection can be insufficient given the diversified urban environments, so we should better look into a holistic profiling (i.e., identifying semantic representations) of \textit{location context}: \textit{location-related contextual information that affects GPS performance}. In particular, location context concerns various scenarios lying between perfect outdoor areas (e.g., large parking areas out of shopping malls) and perfect indoor spaces (e.g., windowless research laboratories surrounded by thick walls). Such a profiling may eventually allow us to fuse various signal sources for achieving ideal urban localization that integrates both indoors and outdoors.

  Such an analysis cannot be performed by accessing only the GPS location indicator distilled from raw GNSS measurements \footnote{Global Navigation Satellite System (GNSS) is a generic term referring to satellite navigation systems such as GPS and Galileo. However, here raw GNSS measurements denote the standard data format shared by all such systems, though we actually obtain (only) GPS-GNSS measurements.} through sophisticated filtering processes, since substantial information loss may have been caused by the significant dimension reduction resulting from convention signal processing pipelines. Therefore, it is reasonable to believe that effectively mining a large volume of these raw measurements could end up with a much more holistic profiling on urban environments than the binary IO-detection. With this initial attempt in profiling, we can not only characterize a location semantically (which an erroneous location indicator likely fails to achieve), but also gain a better understanding (or even correction) of the abnormally high location errors observed in urban areas. Moreover, since new devices~\cite{GNSS-Android} supported by Android 7 have opened up the access to raw GNSS measurements, we can pervasively deploy the profiling service on the off-the-shelf smart devices.

  To this end, we conduct a thorough GNSS data collection in a major metropolitan city, targeting representative scenarios such as sheltered walkways, urban canyons, buildings with an open ground level, and so on. Our aim is to process these data using effective machine learning tools, so as to distill latent representations out of them. These representations, on one hand, are used to construct semantic profiles for characterizing various urban location contexts, and on the other than, they can be used to provide a better accuracy indicator than, for example, the ``blue disk'' offered by Google Maps. One of the major challenges in mining GNSS data is its high dimensionality: each record of GPS-GNSS measurements includes up to 32 satellites, with several tens of fields for each of them. This is significantly different from the time-series or image inputs to machine learning algorithms. Another major challenge lies in the inherent misalignment in feature space over time: the same location can be served by very different satellites at various points in time.\footnote{This is also the reason that we consider profiling location \textit{context}s (rather than merely locations), as other factors beyond pure locations should be taken into consideration.} In order to make sense out of these data, we make the following main contributions:
  \begin{itemize}
    \item We, for the first time, propose to profile location context via mining raw GNSS measurements.
    \item We innovate in the feature engineering pipeline to handle both high-dimensionality and misalignment.
    \item We apply a deep autoencoder to distill compressed representations out of the GNSS data sets.
    \item We develop two preliminary applications of our deep profiling; they deliver semantic representations and better location error estimations, respectively.
    \item We perform extensive data collection and experiments to validate our proposals.
  \end{itemize}

  Note that we are not aiming to deliver yet another optimized GPS~\cite{Kjaergaard-MobiSys2009,Hassanieh-MobiCom2012,Chen-INFOCOM2017}, although it could be a future direction to combine conventional signal processing techniques with machine learning schemes. Instead, our main target is to demonstrate to the community that we have left out a substantial amount of information hidden in GNSS data by deriving only location indicators. And we also deliver some preliminary outcomes of mining such latent information, hoping to evoke more mature applications.

\section{Background and Challenges} \label{sec:rw}
%

  \subsection{GPS-based Localization} 
    As the earliest and most representative GNSS, GPS has been in operation for more than 20 years, and it has been continuously optimized at both higher and lower levels since then. In order to improve energy efficiency, both EnTracked~\cite{Kjaergaard-MobiSys2009} and RAPS~\cite{Paek-MobiSys2010} add adaptivity by leveraging the estimation/prediction from historical GPS readings. Others (e.g., A-Loc~\cite{Lin-MobiSys2010}) also suggest to fuse other sensing information to assist GPS so that to reduce its energy consumption. Thanks to these proposals and related engineering efforts such as A-GPS~\cite{A-GPS}, latest GPS modules in our smartphones have become much more energy efficient than those studied a decade ago~\cite{EnLoc-INFOCOM2009}, especially when only GNSS measurements are retrieved, as will shown in Section~\ref{ssec:energy} using real-life data.

    Optimization at lower level focuses on signal processing and computation. In particular, signal sparsity has been exploited in both \cite{Hassanieh-MobiCom2012} and \cite{Misra-IPSN2014} to speed up computation and thus saving energy. To improve localization accuracy with low-cost receivers, collaboration computation has been introduced by \cite{Hedgecock-SenSys2014} for a scalable network of single-frequency receivers. Further computation reduction can be achieved by cloud offloading~\cite{Liu-SenSys2012}: the GPS receiver collects only a small number of raw GPS readings while moving the location estimation to the cloud. In fact, the raw GPS readings contain a fairly high amount of information, simply summarizing them into 3-axis location indicators using model-based methods almost surely causes information loss. Our attempt to process these raw data using machine learning schemes is the first step towards their more holistic utilization.

  \subsection{Indoor-Outdoor Detection}
    Differentiating indoor and outdoor scenarios has been the first application of the data analytic approach on GPS readings. In fact, earlier approaches of IO-detection~\cite{Zhou-SenSys12,Radu-SenSys14} criticize simple data analytics on the number visible satellites as being slow and energy inefficient. As a result, they apply various machine learning techniques to fuse non-GPS sensor readings: while \cite{Zhou-SenSys12} applies a naive decision tree, \cite{Radu-SenSys14} has tried out several semi-supervised learning techniques. Nevertheless, classical machine learning techniques are too sensitive to feature selection, which can be an obstacle when facing sophisticated data inputs. Recently, SatProbe~\cite{Chen-INFOCOM2017} revisits the number of visible satellites for IO-detection, and it proposes efficient algorithms to extract this feature. However, the access to raw GNSS measurements granted by Android 7 provides users with much more information (e.g., the spatial relation among visible satellites) than just this counter-based feature. Moreover, existing approaches all categorize urban scenarios into 3 classes: outdoor, semi-outdoor, and indoor; this is apparently oversimplifying our real-life location context.

  \subsection{Non-GPS Outdoor Localization}
    Exactly due to the unstable performance of GPS in urban areas, many have considered supplementary solutions~\cite{Cheng-MobiSys2005,Thiagarajan-NSDI2011,WOLoc-INFOCOM2017,CellPos-INFOCOM2018}. Essentially, due to the proliferation of mobile and wireless communications, urban areas have been densely covered by both cellular networks~\cite{Thiagarajan-NSDI2011,CellPos-INFOCOM2018} and WiFi hotspots~\cite{Cheng-MobiSys2005,WOLoc-INFOCOM2017}, and these communication facilities have provided extra signal sources for location estimation. Experience suggests that fusing GPS and non-GPS signal source could yield higher localization accuracy, but this can be achieved only with an adequate profiling of location context.

  \subsection{Mining Mobile Sensing Data}
    Thanks to the latest mobile devices equipped with numerous sensors, we can obtain a huge volume of sensing data, mining which has enabled quite a few applications including notably localization (e.g., \cite{Bahl-INFOCOM00,Youssef-MobiSys2005,Luo-TMC15,WOLoc-INFOCOM2017}) and activity recognition (e.g., \cite{Huynh-UbiComp2008,Matsubara-SIGMOD2014,Parate-MobiSys2014,Liu-MobiCom2016}). Most of these proposals follow a classical machine learning pipeline, starting with empirically selecting a few features and ending with a mapping from these features to certain known patterns. To relieve the reliance on pre-knowledge, Lasagna~\cite{Liu-MobiCom2016} is among the first to apply deep learning technique for activity recognition. Nevertheless, the data inputs considered so far are simply time series (e.g., gyroscope delivers intensity readings on 3 axes), whereas the GNSS data we shall look into have a much more complicated structure, rendering existing proposals inapplicable.  
    \begin{figure*}[!t]
      \centering
        \parbox{\textwidth}{
    \parbox{.16\textwidth}{\center\includegraphics[width=.16\textwidth]{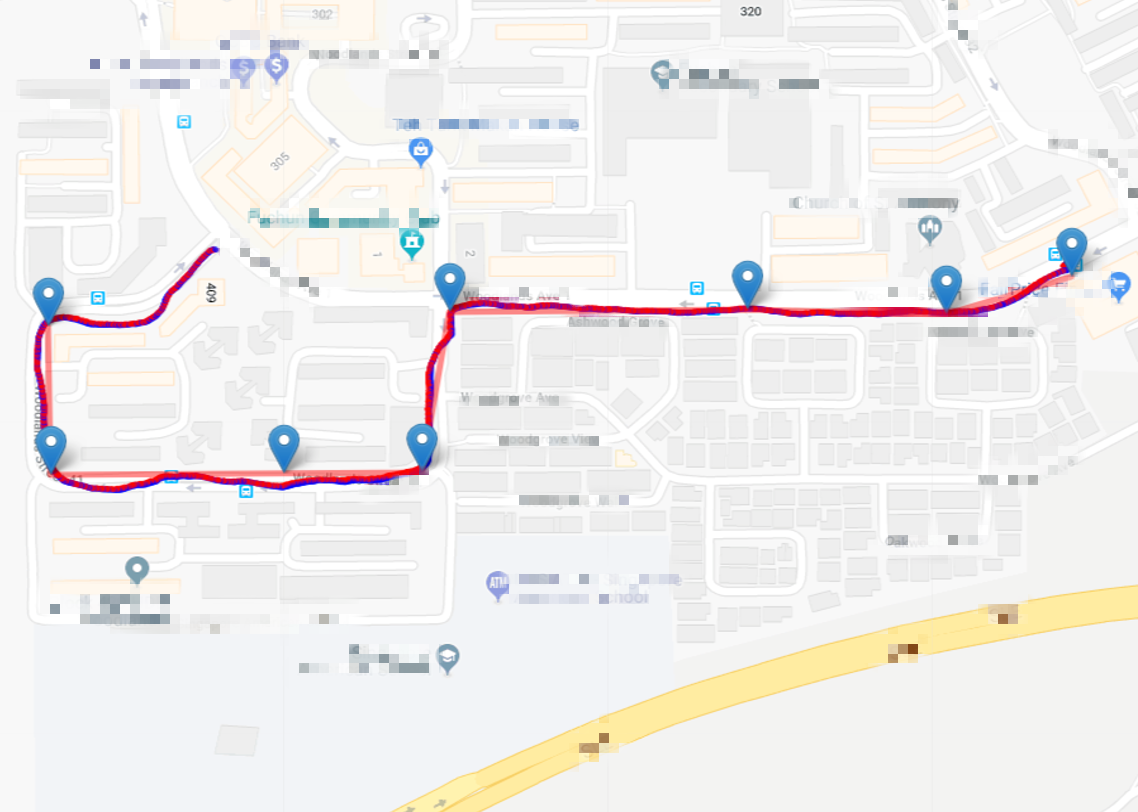}}
        \parbox{.16\textwidth}{\center\includegraphics[width=.16\textwidth]{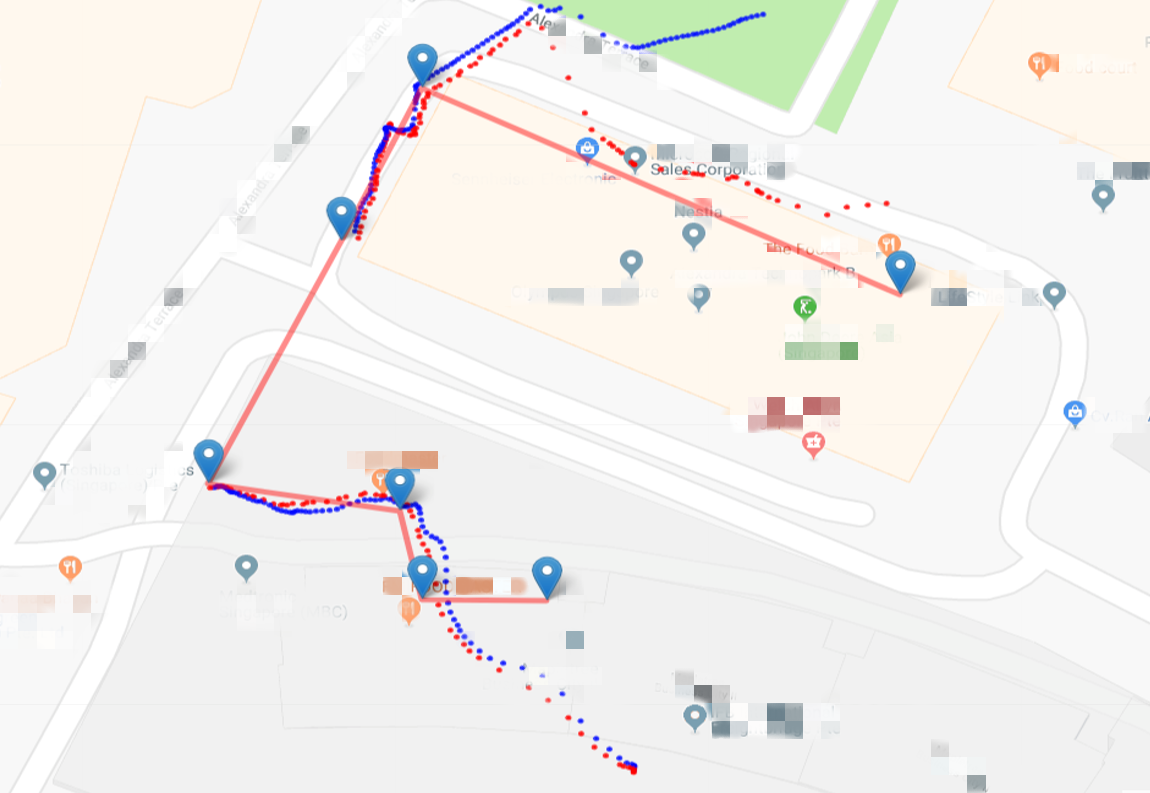}}
            \parbox{.16\textwidth}{\center\includegraphics[width=.16\textwidth]{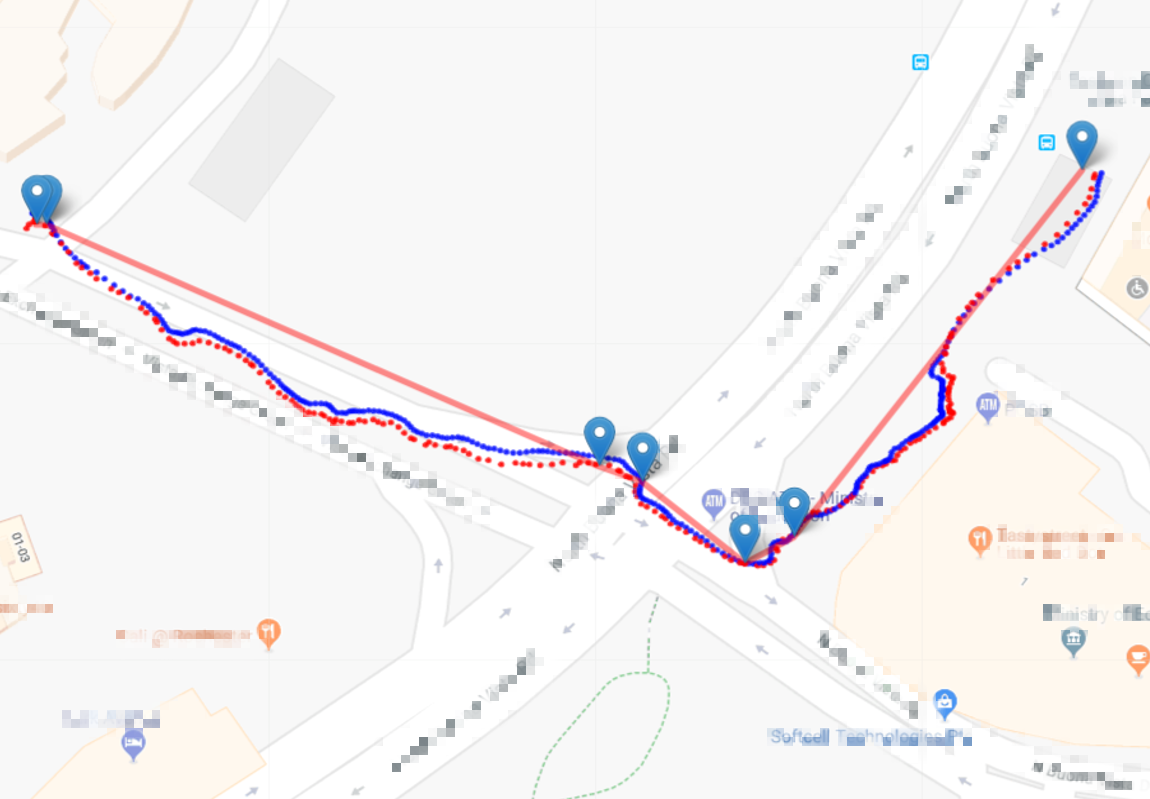}}
            \parbox{.16\textwidth}{\center\includegraphics[width=.16\textwidth]{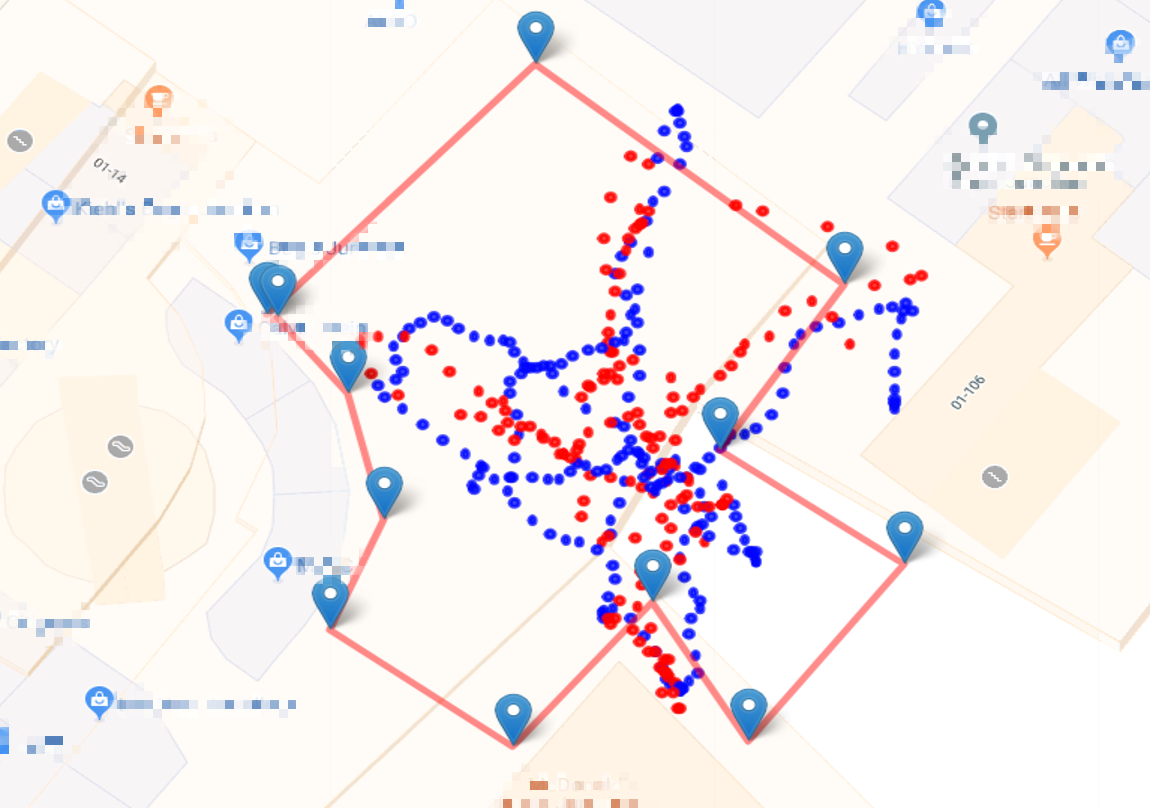}}
            \parbox{.16\textwidth}{\center\includegraphics[width=.16\textwidth]{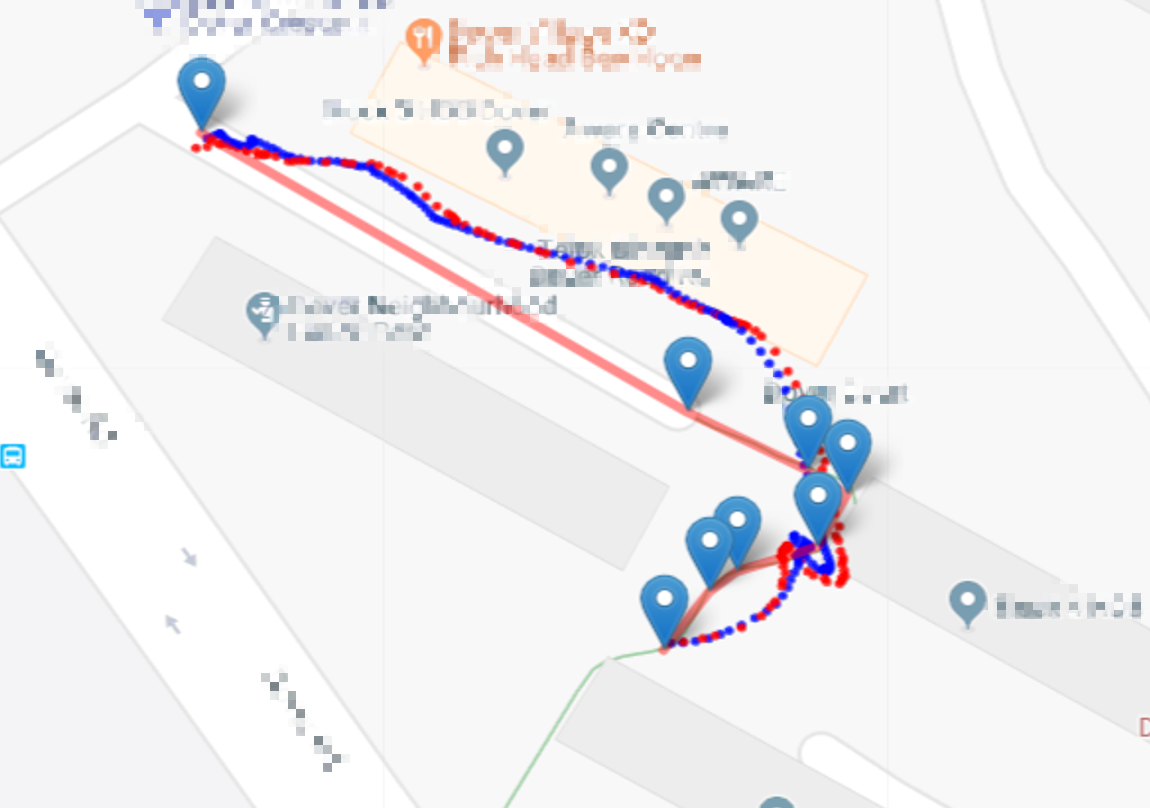}}
            \parbox{.16\textwidth}{\center\includegraphics[width=.16\textwidth]{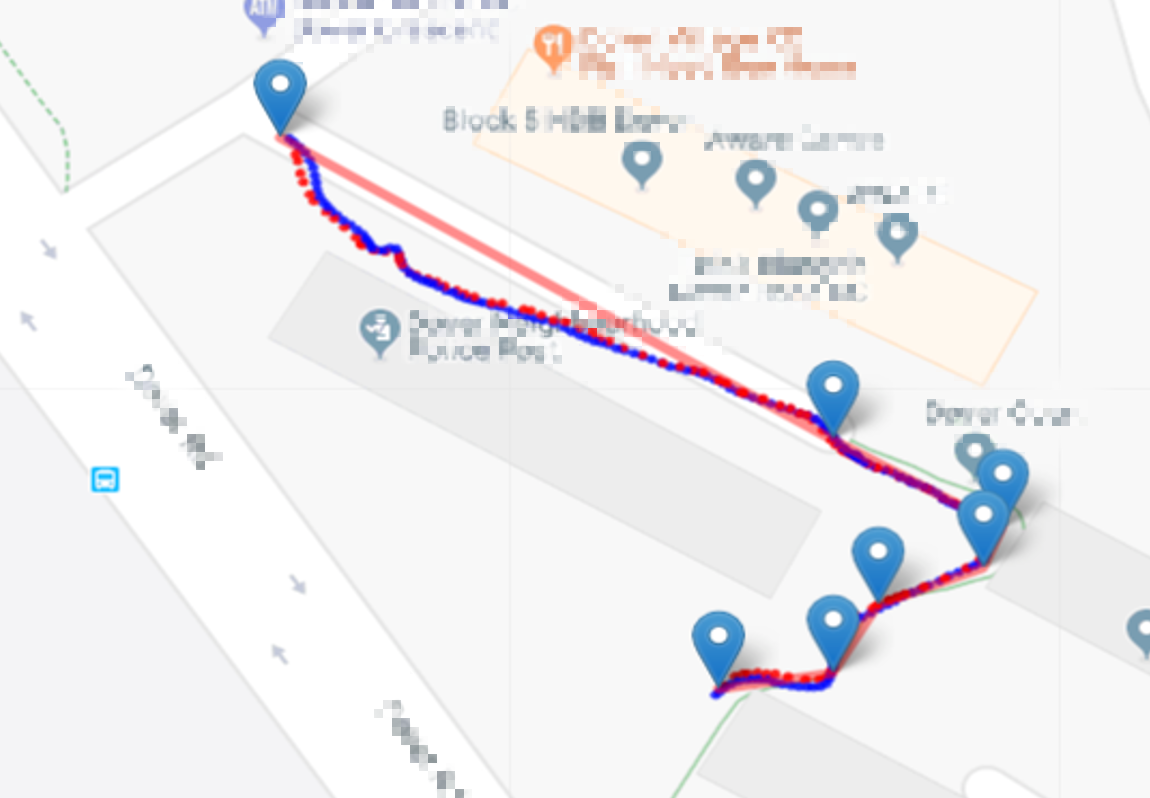}}
                     }\\    
    \parbox{\textwidth}{
          \parbox{.16\textwidth}{\center\includegraphics[width=.16\textwidth]{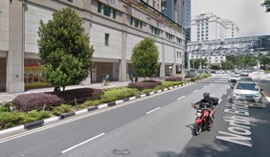}}
        \parbox{.16\textwidth}{\center\includegraphics[width=.16\textwidth]{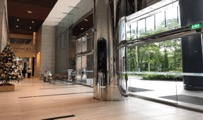}}
            \parbox{.16\textwidth}{\center\includegraphics[width=.16\textwidth]{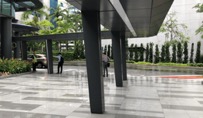}}
            \parbox{.16\textwidth}{\center\includegraphics[width=.16\textwidth]{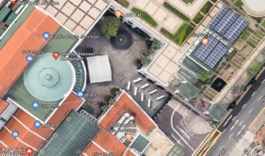}}
            \parbox{.16\textwidth}{\center\includegraphics[width=.16\textwidth]{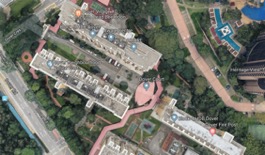}}
            \parbox{.16\textwidth}{\center\includegraphics[width=.16\textwidth]{figure/sat5}}
                     }\\
      \parbox{\textwidth}{
          \parbox{.16\textwidth}{\center\includegraphics[width=.16\textwidth]{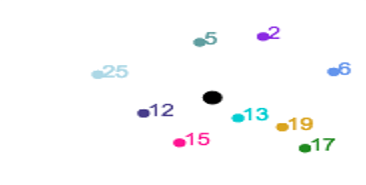}}
        \parbox{.16\textwidth}{\center\includegraphics[width=.16\textwidth]{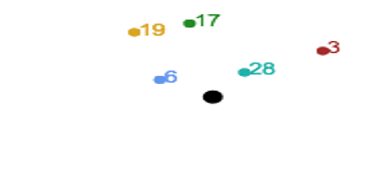}}
            \parbox{.16\textwidth}{\center\includegraphics[width=.16\textwidth]{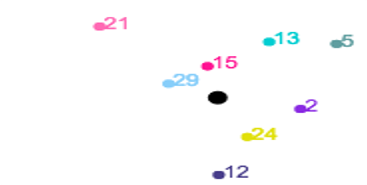}}
            \parbox{.16\textwidth}{\center\includegraphics[width=.16\textwidth]{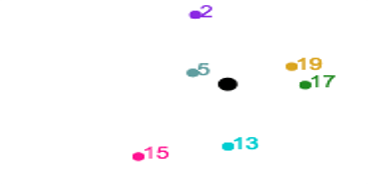}}
            \parbox{.16\textwidth}{\center\includegraphics[width=.16\textwidth]{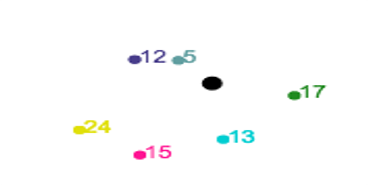}}
            \parbox{.16\textwidth}{\center\includegraphics[width=.16\textwidth]{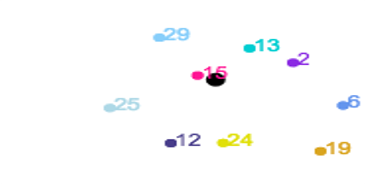}}
                     }\\
        \subfigure[Outdoor along street.] {\label{fig:outdoor}\includegraphics[width=.16\textwidth]{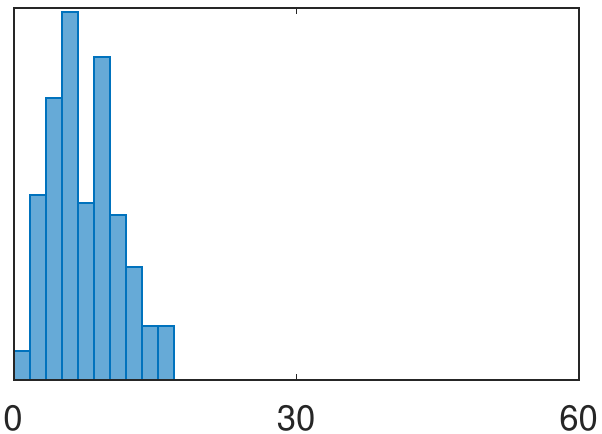}}
        \subfigure[Semi-exposed indoor.] {\label{fig:semi_indoor}\includegraphics[width=.16\textwidth]{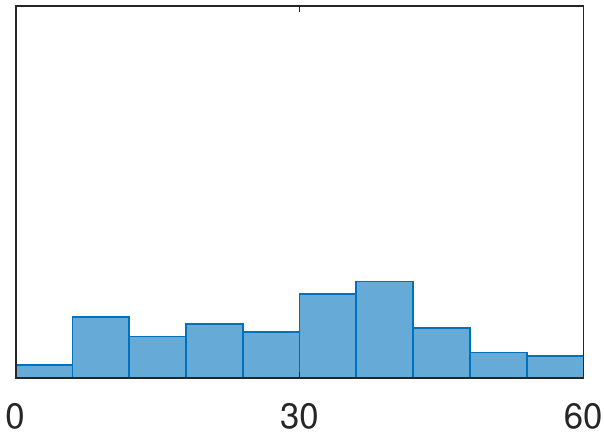}}
        \subfigure[Sheltered street.] {\label{fig:shelter}\includegraphics[width=.16\textwidth]{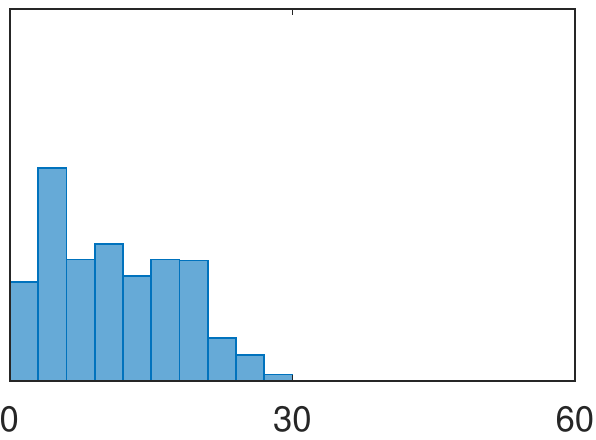}}
        \subfigure[High-rise~enclosing.] {\label{fig:well}\includegraphics[width=.16\textwidth]{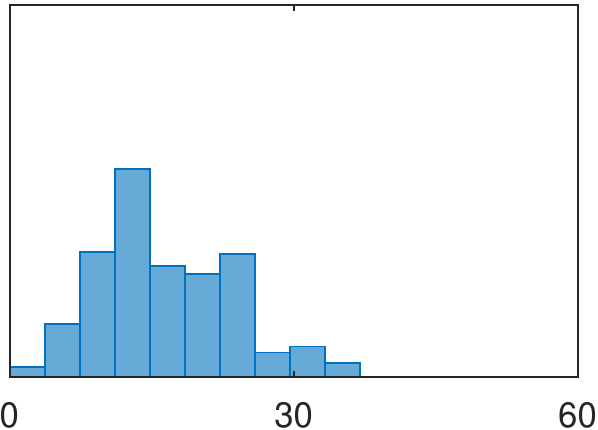}}
        \subfigure[Between buildings.] {\label{fig:between_building}\includegraphics[width=.16\textwidth]{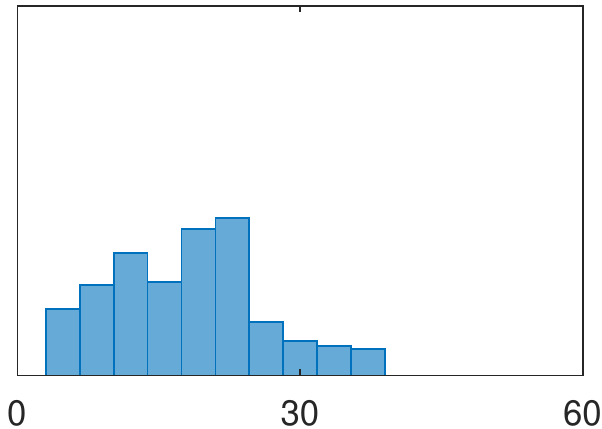}}
        \subfigure[Between buildings.] {\label{fig:between_building_2}\includegraphics[width=.16\textwidth]{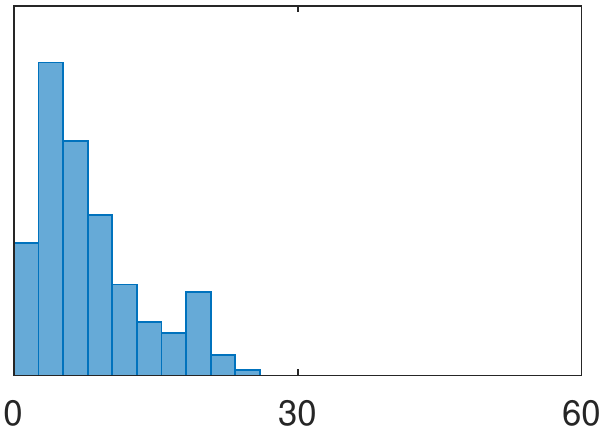}}
     \caption{Preliminary studies on GNSS data. (a) Outdoor long wide street, (b) inside glass building and on void deck, (c) sheltered street, (d) high-rise enclosed area, (e) building surrounded street, and (f) the same as (e) but at a different time. We put ground-truth and GPS traces on Google Maps snapshots in the first row, the corresponding photographic views in the second row, then satellite distribution patterns (with black dot indicating the position of receiver) in the third row, and finally GPS error distributions in the last row.}
     \label{fig:gps_traces}
    \end{figure*}

  \subsection{Challenges in Mining GNSS Data} \label{ssec:challenges}
    While above discussions indicate necessities and potentials of mining GNSS data, several challenges need to be overcome to complete this task. Before diving into the study of GNSS data in Section~\ref{ssec:case}, we briefly hint two major issues here. 
    \begin{itemize}
      \item \textbf{Beyond dimensionality}: the involved data structure is fairly complicated, as shown in \textbf{Table~\ref{tab:fields}} for one satellite. One could deem it as a high dimensional vector and simply stack 32 of them (for all GPS satellites) into a matrix, yet such an input data structure is inefficient: many fields are highly correlated and such redundancy needs to be sorted out before inputting to a machine learning module. In other words, a more effective data representation is hence necessary for a learning module to efficiently capture location contexts under profiling.
      \begin{table*}[h]
      \centering
      \caption{Raw GNSS data structure.} \label{tab:fields}
      \begin{tabular}{|l|l|l|l|}
      \hline
      \multicolumn{2}{|c|}{Binary Fields}                                            & \multicolumn{2}{c|}{Number Fields}                                        \\ \hline
      \multicolumn{1}{|c|}{Field Name}   & \multicolumn{1}{c|}{Description}          & \multicolumn{1}{c|}{Field Name} & \multicolumn{1}{c|}{Description}        \\ \hline
      STATE\_SYMOL\_SYNC                 & is tracking synchronized at symbol level? & ReceivedSvTimeNanos             & the received GNSS satellite time in ns. \\ \hline
      STATE\_TOW\_DECODED                & is time-of-week known?                    & BiasUncertaintyNanos            & the clock’s bias uncertainty in ns.     \\ \hline
      STATE\_BIT\_SYNC                   & is tracking synchronized at bit level?    & PseudorangeRateUncertainty      & rate uncertainty (1-sigma) in m/s.      \\ \hline
      MultipathIndicator                 & has multipath been detected?              & Cn0DbHz                         & the Carrier-to-noise density in dB-Hz.  \\ \hline
      ...                                & ...                                       & ...                             & ...                                     \\ \hline
      \end{tabular}
    \end{table*}
      %

      \item \textbf{Inherent misalignment}: no matter what data structure is used, a proper indexing is a key to align relevant features. Whereas conventional mobile sensing has the sensor IDs as a natural indexing, using satellite IDs for the same purpose can severely misalign the feature: the same location may witness very different satellite coverage at different points in time, while even the same number of visible satellites may form various patterns and thus causing very different localization performance.
    \end{itemize}

\section{From Understanding GNSS Data to Problem Definition} \label{sec:problem_define}
  In order to better understand the problem and its solution feasibility, we have performed extensive studies on the GNSS data and their corresponding location contexts (roughly defined in Section~\ref{sec:intro} but to be elaborated further). In this section, we firstly present a few key observations, which motivates a concrete problem definition, as well as later solution methods.

  \subsection{Case Study of GPS-based Localization Performance} \label{ssec:case}
    During our extensive survey in a city, we have collected the following data for various location scenarios: i) raw GNSS measurements, ii) Android GPS location indicators, iii) ground truth location indicators (human labelled), and iv) Google Maps snapshots. We briefly report a few typical results in \textbf{Figure~\ref{fig:gps_traces}}, where we choose to put the overlay of ii) and iii) onto iv) in the first row, along with the corresponding (photographic) views in the second row, then i) alone in the third row, and finally the errors distribution derived from ii) and iii) in the last row. As it is impossible to exhibit all GNSS data in one figure, we show the derived location patterns for visible satellites, as inspired by constellation-based astronomical localization~\cite{Groth-AJ1986}. The following are a few key observations on these results.
    \begin{itemize}
      \item \textbf{Satellite pattern matters}: a wide street \ref{fig:outdoor} enjoys a balanced satellite pattern and thus an almost perfect (for commercial GPS) average localization error of 5-10 meters, but semi-exposed indoor areas \ref{fig:semi_indoor} (glass-covered building or void deck) cause a biased pattern and larger variations in GPS errors. A balanced pattern obtained for sheltered street \ref{fig:shelter} again leads consistent GPS performance, though the location scenario is quite close to indoor in commonsense. 
      \item \textbf{Correlations between GNSS data and locations exist}: besides the discussions in the previous bullet, high-rise surrounded square or narrow street \ref{fig:well} (scenarios often termed  ``building well'' or ``urban canyon'') lead to a more skewed pattern, as well as a more severe multipath effect (another GNSS field not shown in the figure). Suppose we may properly differentiate \ref{fig:outdoor} from \ref{fig:well} via contextual information inferred by GNSS data, adequate corrections by other complementary signal sources can be put into action. Note that such a differentiation is achieved without deriving GPS location indicators that could indicate locations erroneously.
      \item \textbf{Other GNSS features may further help}: differentiating \ref{fig:outdoor} from \ref{fig:well} may sometimes require more features than the satellite pattern itself. In addition, even when the location remains roughly the same, \ref{fig:between_building} and \ref{fig:between_building_2} exhibit very different GPS tracking performance during two distinct time periods, indicating a change of contextual information beyond pure location. Nevertheless, attributing the context change solely to satellite pattern may not be sufficient; a holistic mining of several GNSS features is necessary. 
    \end{itemize}

    Essentially, our preliminary studies have demonstrated the necessity and benefit of profiling \textit{location context} that mainly involves time-independent location scenario and time-related GNSS features. The importance of satellite pattern (and its independence of satellite IDs) may help us to re-organize the GNSS data to enable an effective mining.

  \subsection{Problem Definition}
    By now, we can clearly define the problem we aim to tackle in this paper. The major task is to profile location context based on raw GNSS measurements. In particular, we consider a large volume of GNSS data records collected by various GPS receivers (of capable smartphones), each \textit{record} containing \textit{sample}s obtained within a time segment $\mathcal{T}$ but from one of $\mathcal{N}$ visible satellites and each sample includes $\mathcal{M}$ fields. Our goal is to first process and pack all the records of dimension $\mathcal{T}\times\mathcal{M}\times\mathcal{N}$ into valid feature map, then we identify a \textit{vectorized semantic representation} through unsupervised learning, aiming at compactly characterizing these seemingly sophisticated data structures. The outcome of this profiling process should yield a representation that satisfies the following criteria to ensure its effectiveness: 
    \begin{enumerate}
      \item It can be further derived and used to estimate an accuracy level of GPS localization performance.
      \item It has specific semantic meaning which can be further clustered or classified to represent various urban location contexts.
      \item Its dimension should be sufficiently reduced, so as to be computational efficient for other applications.
    \end{enumerate}
    Although the ultimate purpose of this profiling is to assist an integrated localization service, we would refrain from being over-ambitious in this preliminary study. Instead, we use the above requirements 1) and 2) to act as two immediate applications, in order to demonstrate the usefulness of our proposed GNSS profiling. 

\section{GNSS-based Location Context Profiling} \label{sec:gnsscp}
  In this section, we present a pipeline to transform raw GNSS measurements into compact data structures adapted to deep learning models. We then engineer an autoencoder model to profile location context based on these data. We further evaluate the effectiveness of this profiling with a supervised task and an unsupervised task.

  \subsection{Feature Engineering Pipeline}\label{ssec: feature_engineering}
    \subsubsection{GNSS Data from Android 7.0}
      Previously, GPS modules handle all the tasks from receiving and processing signals to calculating pseudoranges and thus estimating locations; they only return estimated locations and corresponding accuracy level to front-end applications. Starting from Android 7.0, raw GNSS measurements become available to users via a number of devices~\cite{GNSS-Android}. Each sample of a GNSS measurement includes three general types of fields: i) state indicators that describe clock synchronization, multipath effect and hardware clock continuity, ii) raw measurements, such as carrier frequency, cycles and phases, and iii) computed measurements, such as uncertainty of some raw measurements. A detailed list of fields can be found in \cite{AndroidGNSSAPI}. By invoking the GNSS Raw Measurement API, a sample including all these fields is returned from every visible satellite. In order to facilitate our learning purpose, we regroup them into two categories of features (part of them shown in \textbf{Table~\ref{tab:fields}}) and also derive a few new features. We briefly explain them in the following:
      \begin{itemize}
        \item \textbf{Binary features} take only boolean values and they are related to the clock status, carrier phase validity, and multipath effect. Some of these features may not be directly applicable to the profiling, yet they indicate the validity of other features.
        \item \textbf{Number features} take float number values and mainly include measurements concerning information quality. For example, there are measures concerning SNR and Doppler shift, along with their respective uncertainty fields providing 1-$\sigma$ values. Although these feature are often used to infer localization accuracy given a model-based approach, we intend to use them in a non-parametric manner to avoid losing information.
        \item \textbf{Derived features} are not directly obtained from Android API;  we compute them mainly to facilitate reshaping the input to deep learning modules. In particular, \textit{SvPosition} is the position (under spherical coordinate system) of a satellite when a GNSS sample is received, derived using the GPS ephemeris~\cite{gps_data} and the receiving timestamps. 
      \end{itemize}

    \subsubsection{Satellite Position based Feature Packing} \label{sssec:packing}
      \begin{figure}[htb]
      \centering
      \includegraphics[width=.8\columnwidth]{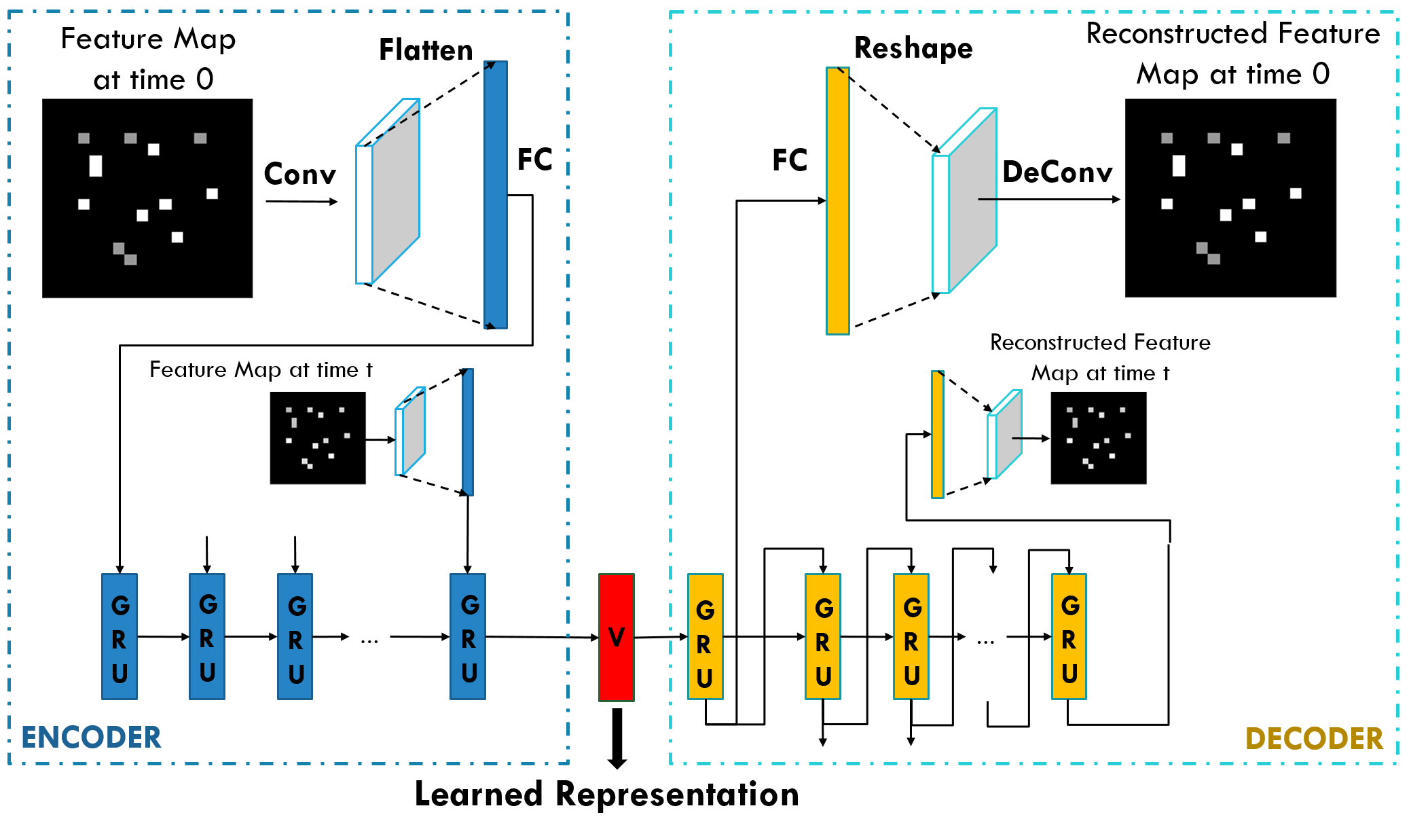}
      \caption{Autoencoder structure.} 
      \label{fig:auto_encoder}
      \end{figure}
\begin{figure}[!t]
      \centering
      \parbox{\textwidth}{
            \parbox{.24\textwidth}{\center\includegraphics[width=.24\textwidth]{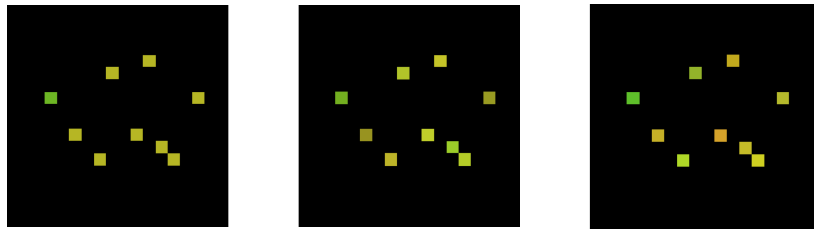}}
          \parbox{.24\textwidth}{\center\includegraphics[width=.24\textwidth]{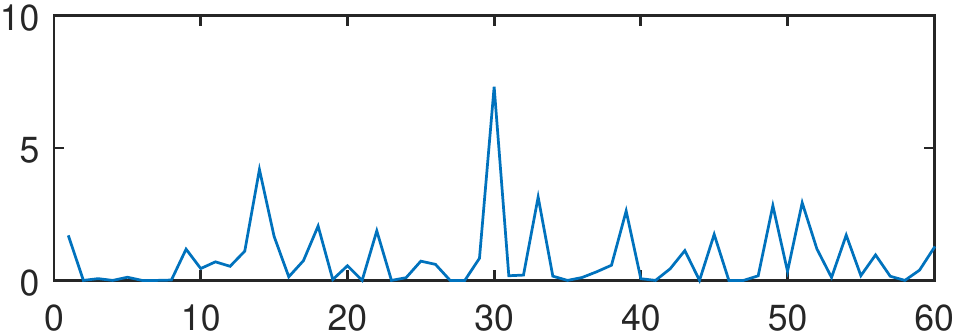}}
                       }\\
       \parbox{\textwidth}{
            \parbox{.24\textwidth}{\center\includegraphics[width=.24\textwidth]{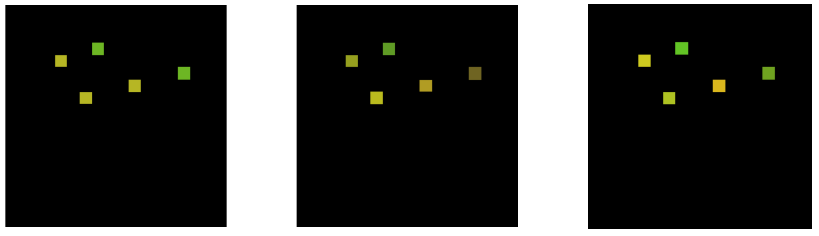}}
          \parbox{.24\textwidth}{\center\includegraphics[width=.24\textwidth]{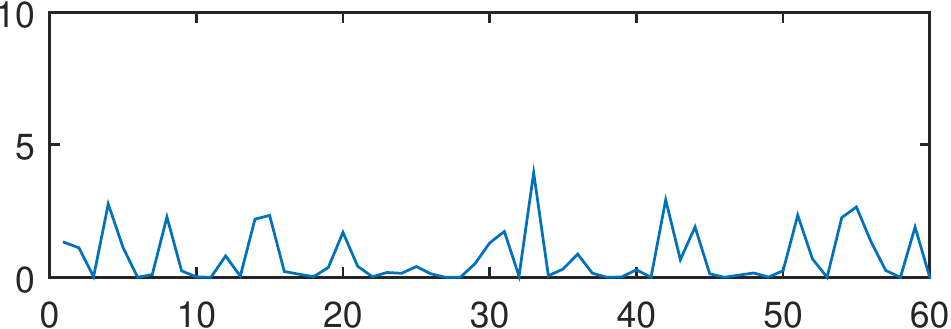}}
                       }\\
      \parbox{\textwidth}{
            \parbox{.24\textwidth}{\center\includegraphics[width=.24\textwidth]{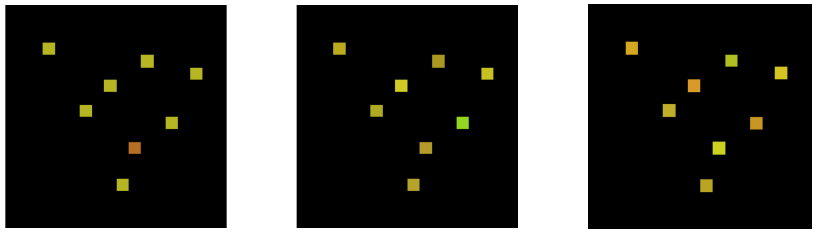}}
          \parbox{.24\textwidth}{\center\includegraphics[width=.24\textwidth]{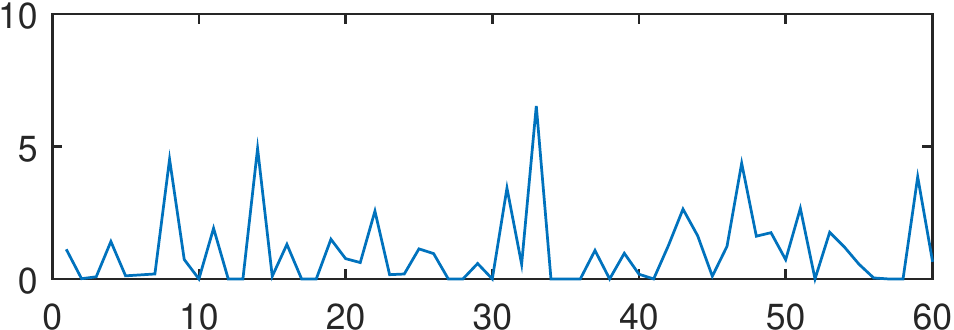}}
                       }\\
       \parbox{\textwidth}{
            \parbox{.24\textwidth}{\center\includegraphics[width=.24\textwidth]{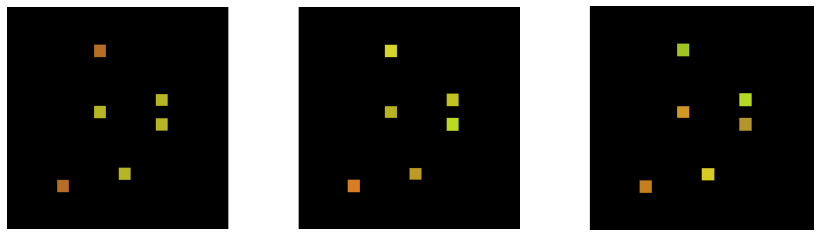}}
          \parbox{.24\textwidth}{\center\includegraphics[width=.24\textwidth]{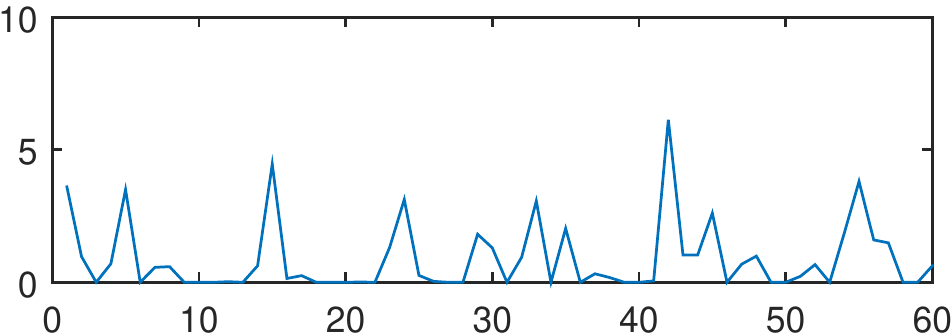}}
                       }\\
       \parbox{\textwidth}{
            \parbox{.24\textwidth}{\center\includegraphics[width=.24\textwidth]{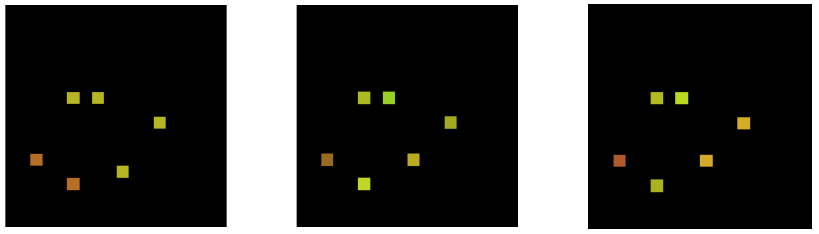}}
          \parbox{.24\textwidth}{\center\includegraphics[width=.24\textwidth]{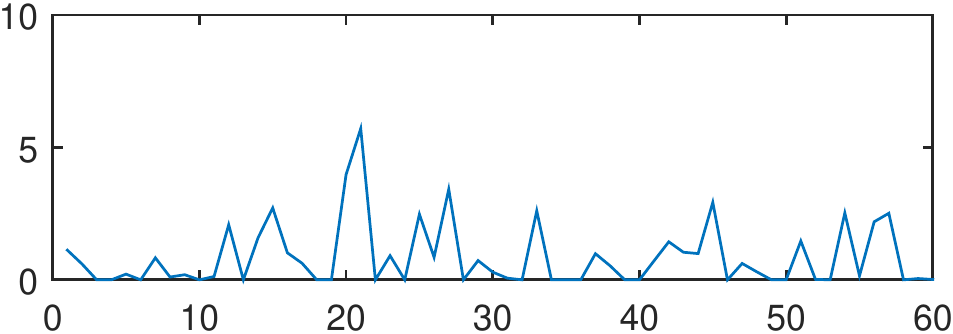}}
                       }\\
       \parbox{\textwidth}{
            \parbox{.24\textwidth}{\center\includegraphics[width=.24\textwidth]{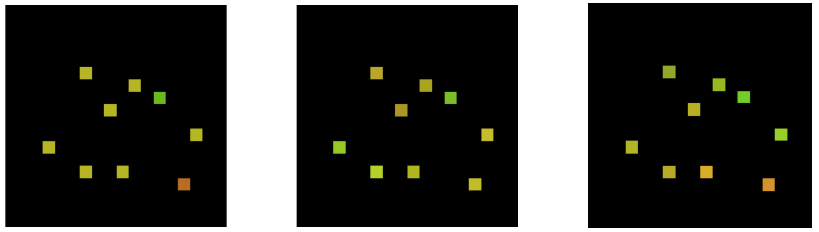}}
          \parbox{.24\textwidth}{\center\includegraphics[width=.24\textwidth]{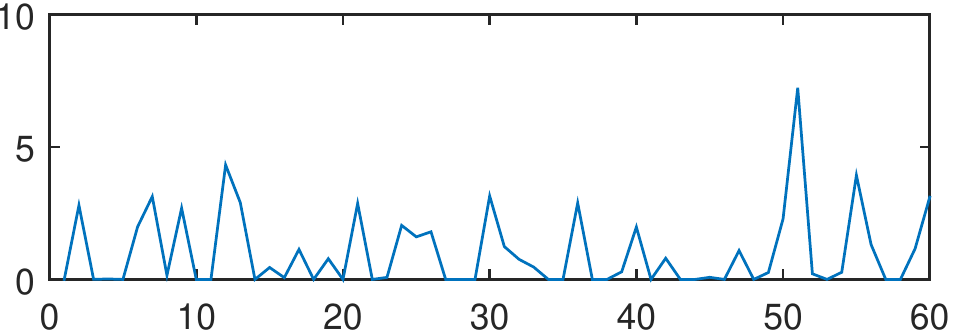}}
                       } \vspace{-2ex}
      \subfigure[Snapshots at $\tau=1,2,3$.] {\label{fig:original}\includegraphics[width=.48\columnwidth,height=1pt]{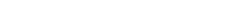}}
      \subfigure[Representation $\mathbf{v}$.]{\label{fig:log}\includegraphics[width=.48\columnwidth,height=1pt]{figure/empty}}
      \caption{Snapshots and corresponding representations.}
      \label{fig:sv_feature_img}
      \end{figure}

      As we have elaborated in Section~\ref{ssec:challenges}, packing samples for all satellites into one compact and meaningful representation is highly non-trivial. Fortunately, the preliminary studies in Section~\ref{ssec:case} indicate that the satellite pattern might be a good way of indexing the complex data structure. Further inspired by the fact that most deep learning models take images (i.e., 2D pixel maps) as common input, we design an pixel-position based packing mechanism. Essentially, we represent each feature of all visible satellites as a \textit{snapshot}, i.e., an image with certain pixels representing the satellites and the pixels taking their values from the features. Each snapshot is centered at the receiver's location and the relative satellite positions derived from \textit{SvPosition} are normalized to determine the corresponding pixel positions.
      As a result, each sample (for all satellites) contains $\mathcal{M}$ snapshots and a record has $\mathcal{T}\times\mathcal{M}$ of them. We could further pack the temporal dimension into snapshot, but that would require a much higher resolution and may also lose track of temporal correlation. We showcase a few snapshot samples in \textbf{Figure~\ref{fig:original}}.

      To ensure the scale and trends of features are consistent, normalization and re-ranking are done before packing them into snapshots. The normalization phase transforms feature values within range $[0,1]$, and it should produce a ``scattered'' distribution for all features. In particular, most uncertainty features follow a log-normal distribution, so we convert them logarithmically to obtain a quasi-normal distribution. The re-ranking phase makes sure all the features at each pixel has a consistent trend in physical meaning. Since pixel values are in greyscale (i.e., non-negative integers) and zero value indicates no valid data presented, a smaller value should correspond to a higher uncertainty. Therefore, all the uncertainty and ambiguity features are reversely ordered to maintain consistency.
  
  \vspace{-6ex}
  \subsection{Location Context Autoencoder}\label{ssec:auto_encoder}
    To learn a compact representation from the packed features (i.e., image sequences), we naturally choose an autoencoder for embedding~\cite{image_autoencoder}, and the compact representations are extracted from the central layer. Sequence inputs are handled by incorporating recurrent neural network (RNN) as the fundamental structure~\cite{sequence-to-sequence}. Given a sequence of inputs and a desired output sequence, the goal of the RNN is to estimate the conditional probability of output sequence given the input sequence. Incorporating the two structures, we design our autoencoder as a sequence-to-sequence structure with images as input. Both the encoder and the decoder are implemented using gated recurrent unit (GRU) cells. As shown in \textbf{Figure~\ref{fig:auto_encoder}}, given the snapshots from time 0 to $\mathcal{T}$, denoted as $\mathbf{x} = (x_0,\cdots,x_\tau\cdots,x_\mathcal{T})$, where $x_\tau\in\mathbb{R}^{n_i\times n_i\times \mathcal{M}}$, $n_i$ is the size of a snapshot, and $\mathcal{M}$ is number of GNSS features, we want to get a compact representation $\mathbf{v} \in \mathbb{R}^\ell$ with $\ell$ being the latent dimension of GRU (shown as red in \textbf{Figure~\ref{fig:auto_encoder}}). The encoder and decoder are defined as transitions $\Theta$ and $\Phi$. Here, $\Theta: \mathbf{x} \rightarrow \mathbf{v}$, $\Phi: \mathbf{v} \rightarrow \mathbf{x}'$, where $\mathbf{x}'$ denotes the reconstructed sequence of $\mathbf{x}$ with the same dimension. The training is conducted with back-propagation algorithm to find transition $\Theta^*$ and $\Phi^*$ based on the following objective:
    \begin{equation*}
      \{\Theta^*,\Phi^*\} = \arg\min_{\Theta,\Phi} \sum_{\tau=0}^{\mathcal{T}}\|x_\tau-x'_\tau\|_2^2.
    \end{equation*}
 
    Encoder first applies a convolution layer with filter size $(m_f,m_f,\ell)$ to each $x_t$ and flattens it into a vector as input for GRU cell. At decoder side, the context vector $\mathbf{v}$ is set as the hidden state, the output generated by RNN is reshaped and applied a deconvolutional layer with the same filter size as applied at the encoder side to obtain the final output $x'_t$. Parameters setting for $m_f$, $\ell$ and $\mathcal{T}$ will be studied in Section~\ref{sssc: parameter}. \textbf{Figure~\ref{fig:sv_feature_img}} shows examples of snapshots and generated representations, based on the 6 scenarios in \textbf{Figure~\ref{fig:gps_traces}}. Due to space limit, we only visualize the first 3 steps in a sequence of length 5, and select 3 features for RGB channels of images.

  \subsection{Applications of Location Context Profiling} \label{ssec:application}
    \subsubsection{Localization Error Estimator}\label{ssec:error_estimator}
      As shown in Section~\ref{ssec:case}, different contexts can result in various error distributions. So the profiling vector $\mathbf{v}$ can be used to estimate localization performance. Such information can yield a confidence level of GPS localization similar to the ``blue disk'' provided by Google Maps. At any time $t$, given the representation $\mathbf{v}_t$ generated from GNSS measurements during previous $\mathcal{T}$ time steps, an error estimation function $\Lambda$ and true GPS error $e_t$, and the predicted error $\hat{e}_t = \Lambda(\mathbf{v}_t)$, we aim to find $\Lambda^* = \arg\min_\Lambda\|e_t-\hat{e_t}\|_2^2$. Using this objective as the loss function, we design a 4-layer feed-forward deep neural network structure with a single neuron at the last layer as our error estimator model. We denote it by DNN-R hereafter.

      We also implement two other deep learning structures, namely DNN-C and RNN, to compare with DNN-R. DNN-C converts the regression problem into a classification problem, in which we adopt a classification layer before the regression neuron layer, and make the final layer as the expectation on the class distribution. RNN implements GRU cells to include estimation in previous steps; it adapts to the filtering effect inside GPS that results in relatively smooth changes in location estimations and corresponding errors. Detailed settings are presented in Section~\ref{ssec:eva_error}.

    \subsubsection{Context Semantic Analysis}\label{sssec:semantic_analysis}
      \begin{figure}[htb!]
      \centering
      \subfigure[Detailed context for trace 1.] {\label{fig:lower1}\includegraphics[width=.23\textwidth]{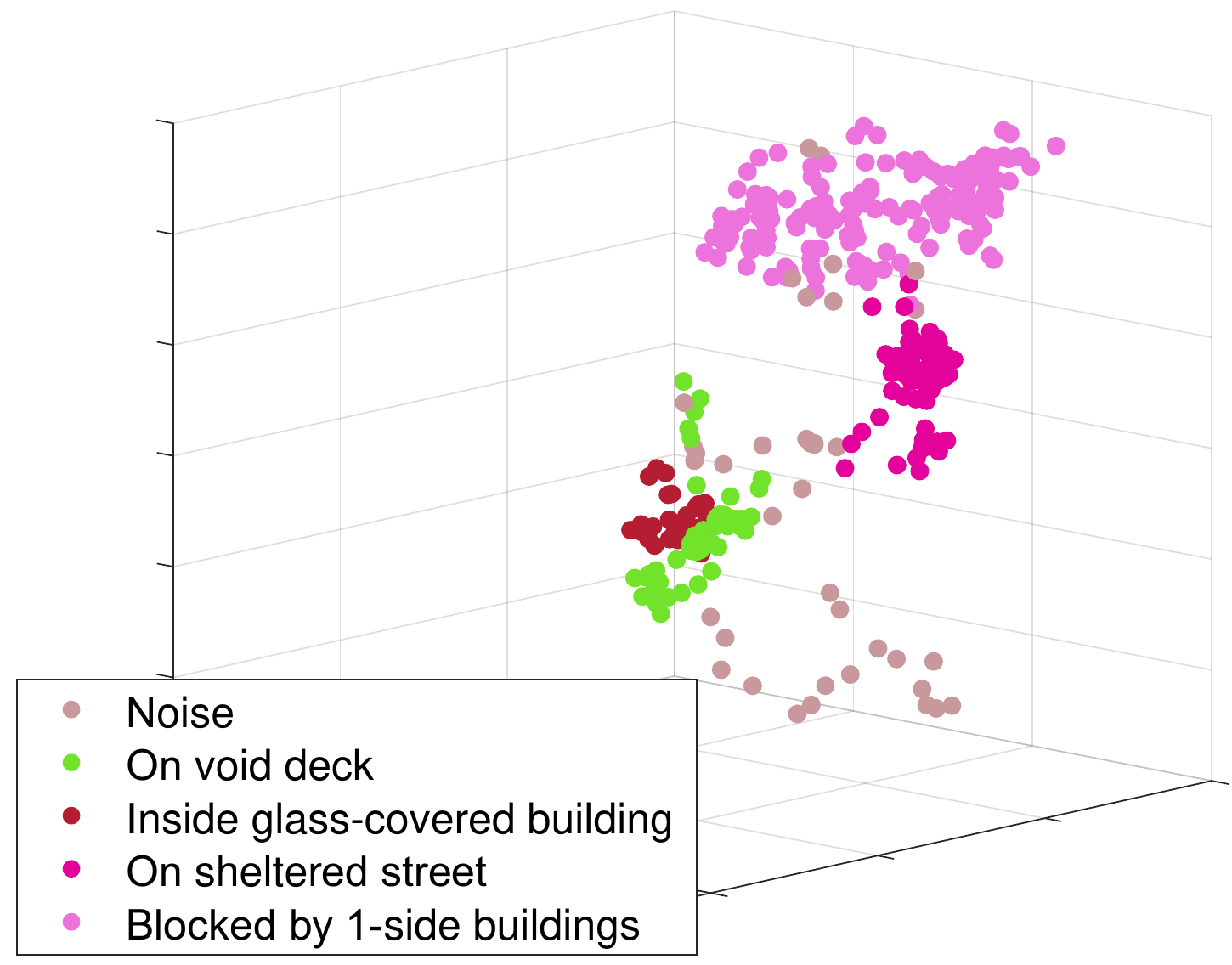}}
      \subfigure[Detailed context for trace 2.] 
      {\label{fig:lower2}\includegraphics[width=.23\textwidth]
      {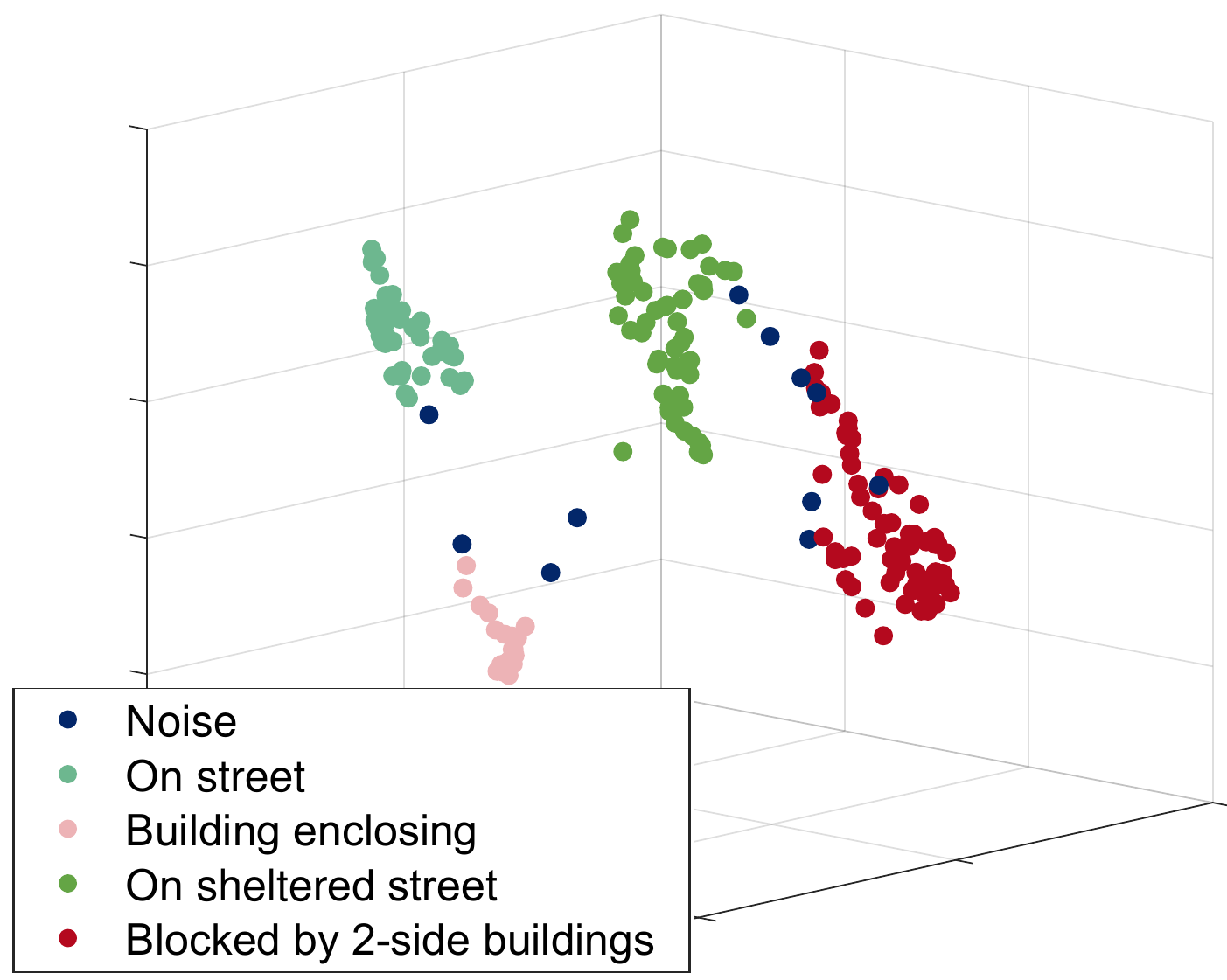}}
      \vspace{3ex}
      \subfigure[General Context.]
      {\label{fig:upper}\includegraphics[width=.3\textwidth]{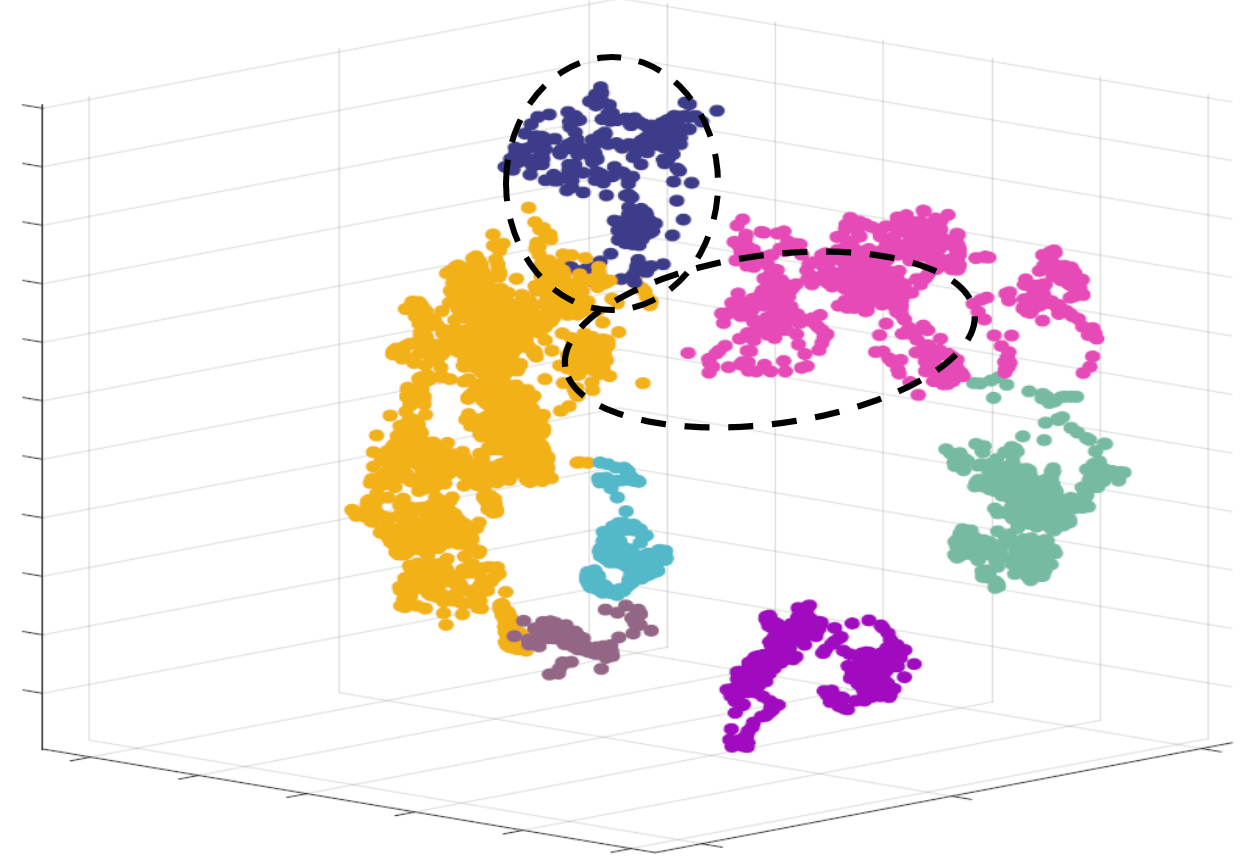}}
      \caption{Context semantic analysis.}
      \label{fig:semantic_analysis}
      \end{figure}
      
      Our studies in Section~\ref{ssec:case} suggest that GNSS measurements can represent the context of environments, which can be used to trigger sensor fusion for better localization. We aim to create a context database containing labelled training representations and allow a new representation to query for a context label. Since our representation is learned from records, a semantic label is assigned to a trip segment (of length $\mathcal{T}$) rather than a spot sample. Moreover, manually labelling representations can be meaningless given the great variety in human perception, so we adopt a bottom-up clustering mechanism to build up the labelled database. Each training representation has 2-level of labels. 
      \begin{itemize}
        \item \textbf{Lower-level clustering}: For a set of representations, we use DBSCAN to cluster them into small clusters, assign each cluster with a predefined label as the \textit{detailed context} and calculate the center of the cluster. Detected noises are just labelled as ``noise''. Lower level clustering captures information of scenario such as ``on street'', ``blocked by 1-side building'', ``building enclosing'', and ``on void deck''. \textbf{Figures~\ref{fig:lower1}} and \textbf{\ref{fig:lower2}} demonstrate two traces after lower-level clustering.
        \item \textbf{Higher-level clustering}: Cluster centers and ``noises'' from lower-level clustering are further combined into upper level of clusters, and assigned a unified label as \textit{general context}. Higher-level clustering captures general information such as satellite patterns. The noise vector detected from upper level of clustering are treated as a separated cluster since it may capture an infrequent satellite pattern at the moment, but can be further clustered into other groups when the database is enlarged. To illustrate the correspondence with lower-level clustering, \textbf{Figure~\ref{fig:upper}} shows the representations with the color label of upper level clusters. 
      \end{itemize}

      At query stage, each representation will be firstly matched to the nearest upper-level clusters to determine its general context, and then compared with all representations inside the cluster to find its k nearest neighbours and assign the final label by majority voting. We set a threshold for finding the kNN so that queries with nearest distance larger than the threshold will not be assigned with any label, because we expect it as a new context representation that is not included in the training database. Note that all the training procedures presented in Section~\ref{sec:gnsscp} are performed offline, making the online computations (basic algebraic operations) in smartphones very light-weight.
      %
      %

\section{Evaluation} \label{sec:eva}
  In this section, we firstly provide details on our system implementation, data preparation, and model parameter settings, then we evaluate the effectiveness of context profiling by examining the performance of the two derived applications. 

  \subsection{System Implementation and Data Preparations}
    We develop applications in Android with Java, using 4 smartphones for data collection: Huawei P10, P10 Plus, Mate 9, and Samsung S8. The entire data processing and model training pipeline is implemented in Python, and set up on a PC with 2.6 GHz Intel Core i7 CPU and a 16 GB 1600 MHz DDR3 memory.

    \subsubsection{Data Collection}\label{sssc:data_collection}
      Our Android application collects raw GNSS measurements and locations returned by GPS. The GPS location indicator has 2 sources in Android: AndroidLoc obtained through Android LocationManager API~\cite{AndroidLocationApi} and GoogleLoc provided by Google FusedLocationProviderApi~\cite{GoogleFusedApi}. We record location and its accuracy (associated with Location class) for both sources at a frequency of 1\!~Hz. The ground truth is manually labelled by marking critical points along the way, and the locations between the critical points are linearly interpolated by time assuming users moving at a constant speed. We collect 30 traces at different times covering about 9,000 typical location points in the urban area, roughly summarized into the following 9 scenarios: 
      \begin{table}[H]
      \centering
      \scriptsize
      \begin{tabular}{lll}
      On Street                   & On void deck        & Blocked by 1-side buildings   \\
      Blocked by 2-side buildings & Building enclossing & Inside glass-covered building \\
      On sheltered street         & In tunnel           & Inside room with heavy walls 
      \end{tabular}
      \end{table}
      %
      %
      We randomly select 20 traces for the training purpose and make the remaining traces as testing traces, resulting in a training dataset with about 6600 location points and a testing dataset with about 2300 location points.

    \subsubsection{Data Preprocessing}
      We match the corresponding raw GNSS measurements, GPS locations, accuracy (from two sources), and ground truth locations by their timestamps. We further calculate the \textit{true GPS localization error}s as the great circle distance between GPS locations and their respective ground truth locations. As explained in Section~\ref{sssec:packing}, we use SvPosition as index to pack features into snapshot. We set the snapshot (image) resolution as $18 \times 18$. The following 18 features are chosen; their details can be found in \cite{AndroidGNSSAPI}: 
      \begin{table}[H]
      \centering
      \scriptsize
      \begin{tabular}{|l|l|}
      \!\!\!\!ADR\_STATE\_CYCLE\_SLIP     & \!\!\!\!HardwareClockDiscontinuityCount \vspace{.5ex} \\
      \!\!\!\!ADR\_STATE\_RESET               & \!\!\!\!BiasUncertaintyNanos            \vspace{.5ex} \\ 
      \!\!\!\!ADR\_STATE\_VALID               & \!\!\!\!ReceivedSvTimeUncertaintyNanos  \vspace{.5ex} \\
      \!\!\!\!STATE\_SYMBOL\_SYNC             & \!\!\!\!Cn0DbHz               \vspace{.5ex}\\
      \!\!\!\!STATE\_MSEC\_AMBIGUOUS          & \!\!\!\!PseudorangeRateMetersPerSecond \vspace{.5ex} \\ 
      \!\!\!\!STATE\_TOW\_DECODED             & \!\!\!\!PseudorangeRateUncertaintyMetersPerSecond \vspace{.5ex} \\ 
      \!\!\!\!STATE\_SUBFRAME\_SYNC           & \!\!\!\!AccumulatedDeltaRangeUncertaintyMeters  \vspace{.5ex}\\ 
      \!\!\!\!STATE\_BIT\_SYNC                & \!\!\!\!DeltaRange \vspace{.5ex} \\
      \!\!\!\!STATE\_CODE\_LOCK               & \!\!\!\!MultipathIndicator
      \end{tabular}
      \end{table}

      We note that our current experiments are rather preliminary: the data traces collected are far from comprehensive and the GNSS features are selected more on an intuitive basis. Nevertheless, we believe they are sufficient to demonstrate the effectiveness of mining GNSS measurements and to provoke further researches on related topics. In particular, crowdsensing (e.g., used to gather Wi-Fi data for outdoor localization~\cite{WOLoc-INFOCOM2017}) can help to significantly enlarge the coverage of collecting GNSS measurements.

  \subsection{Autoencoder Model Training}
    \subsubsection{Parameter Setting} \label{sssc: parameter}
\begin{figure}[htb!]
      \centering
      \subfigure[Latent dimension.] {\label{fig:latent_dim}\includegraphics[width=.23\textwidth]{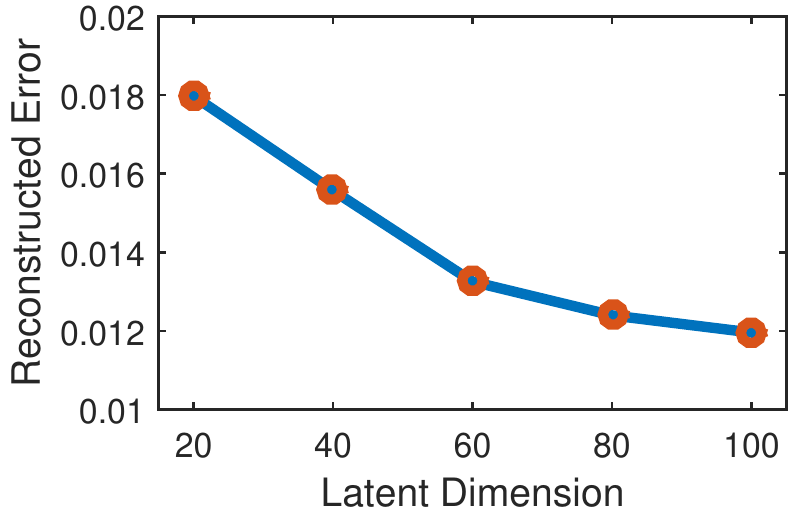}}
      \subfigure[Convoluation filter size.]{\label{fig:filter_size}\includegraphics[width=.23\textwidth]{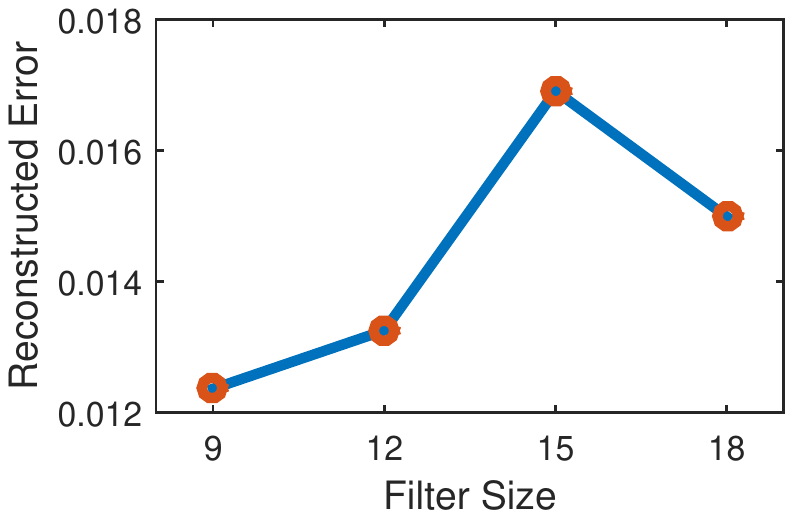}}
      \caption{Autoencoder parameters vs reconstruction error.}
      \label{fig:auto_encoder_para}
      \end{figure}

      As presented in Section \ref{ssec:auto_encoder}, there are 3 groups of hyperparameters to be determined: i) input sequence steps for sequence-to-sequence model, ii) latent dimension for RNN GRU cells, and iii) filter size for convolutional layer. For time steps, we set it to 5 thus the model includes measurements from past 5 seconds. We set such a relatively small number as we intend to make the context relatively stable within the period so that the resulted representation can better characterized a stable context. As the loss for our model is the reconstruction error, we evaluate the effect of these parameters through reconstruction error as shown in \textbf{Figure~\ref{fig:auto_encoder_para}}. The latent dimension directly affects the dimension of our representation. There is a trade-off in determining the kernel number. On one hand, a larger latent dimension allows representations to encode more information. On the other hand, a larger latent dimension will bring extra cost for storage, model training and application at later stage. To achieve a balance, we select 60 when the error starts to drop down slowly. For convolutional filter size, we set it as $(12,12)$ so that each patch can capture sufficient image while achieving an acceptable reconstruction performance. We leverage Adam Optimizer and ReLU activation function for training.

    \subsubsection{Comparison with Other Embedding Mechanism} \label{sssc:crbm}
      As a comparison, we also build a Convolutional Restricted Boltzmann Machine (CRBM) as presented in \cite{Liu-MobiCom2016}. To incorporate the same structure, we treat satellites identifier as their dimensions (D) and various features as the channel (C). Since the feature packing is different, we omit the comparison in reconstruction error but compare the effectiveness of representation next based on the applications discussed in Section~\ref{ssec:application}.

    \subsection{Localization Error Estimator}\label{ssec:eva_error}
      To evaluate the effectiveness of representation, we firstly examine its performance in estimating localization error. We build 3 types of location error estimators as described in Section~\ref{ssec:error_estimator}: DNN-R, DNN-C, RNN; they share the same loss function as the mean squared error between regression result and true GPS error. DNN-R consists of 4 layers with 32, 16, 8, 1 neurons respectively and takes ReLU as the activation function. DNN-C has a similar structure as DNN-R with 4 layers, but the layer with 8 neurons has an additional activation layer of soft-max aiming to output the error range classification for 8 classes. The errors are divided into 8 bins, and each has a size (error range) of 8 meters. The single neuron on the last layer takes the expectation on all error bins. RNN shares a similar structure as DNN-R except that the first 3 layers are recurrent neural layers with GRU cells.
      \begin{figure}[t]
      \centering
      \subfigure[Mean error.] {\label{fig:mean_loc_error}\includegraphics[width=.23\textwidth]{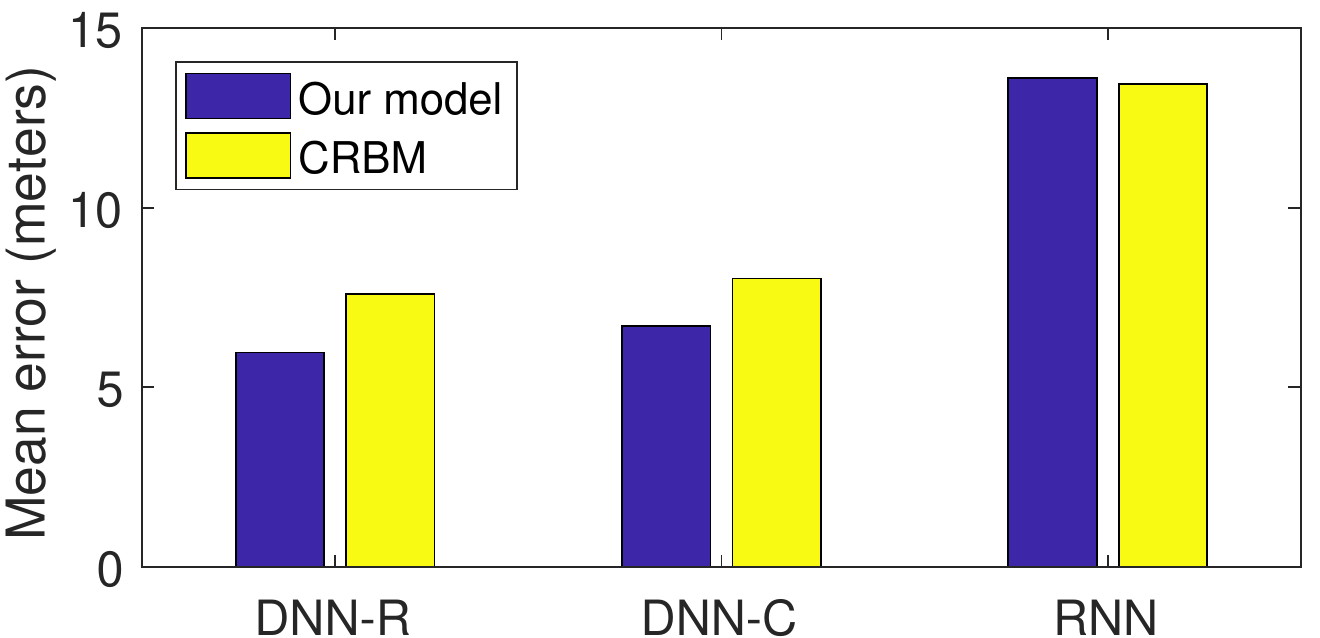}}
      \subfigure[Median error.]{\label{fig:median_loc_error}\includegraphics[width=.23\textwidth]{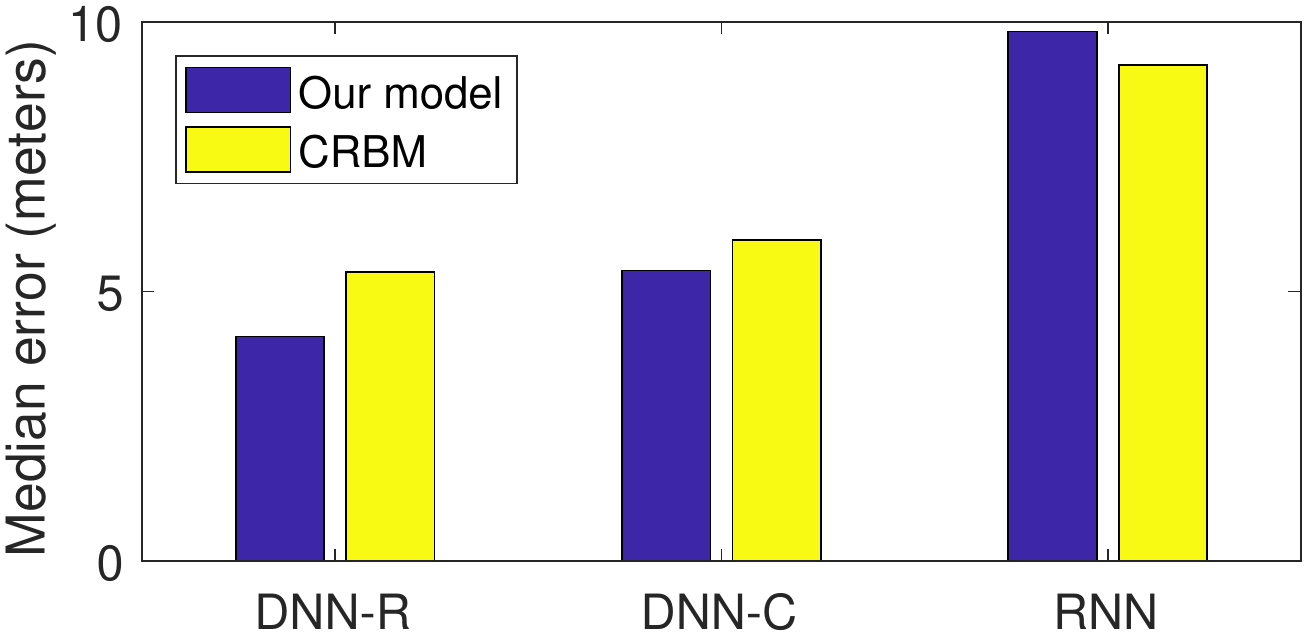}}
      \caption{Deviations from true localization error.} 
      \label{fig:error_comparison}
      \end{figure}

\begin{figure}[b]
      \centering
      \parbox{\textwidth}{
            \parbox{.24\textwidth}{\center\includegraphics[width=.23\textwidth]{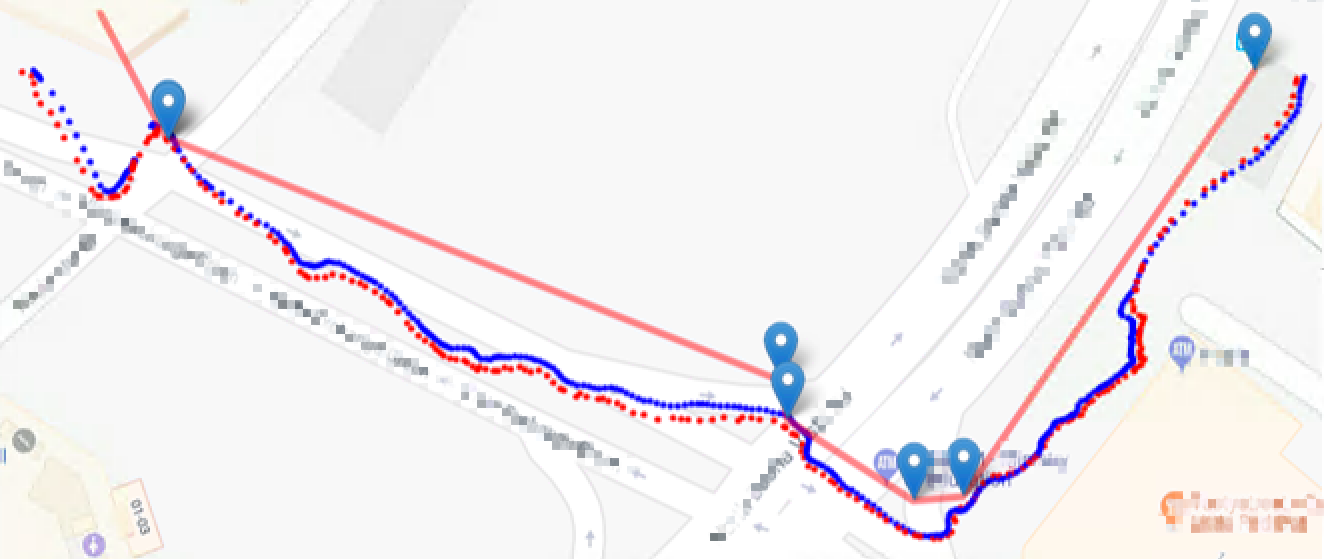}}
          \parbox{.24\textwidth}{\center\includegraphics[width=.23\textwidth]{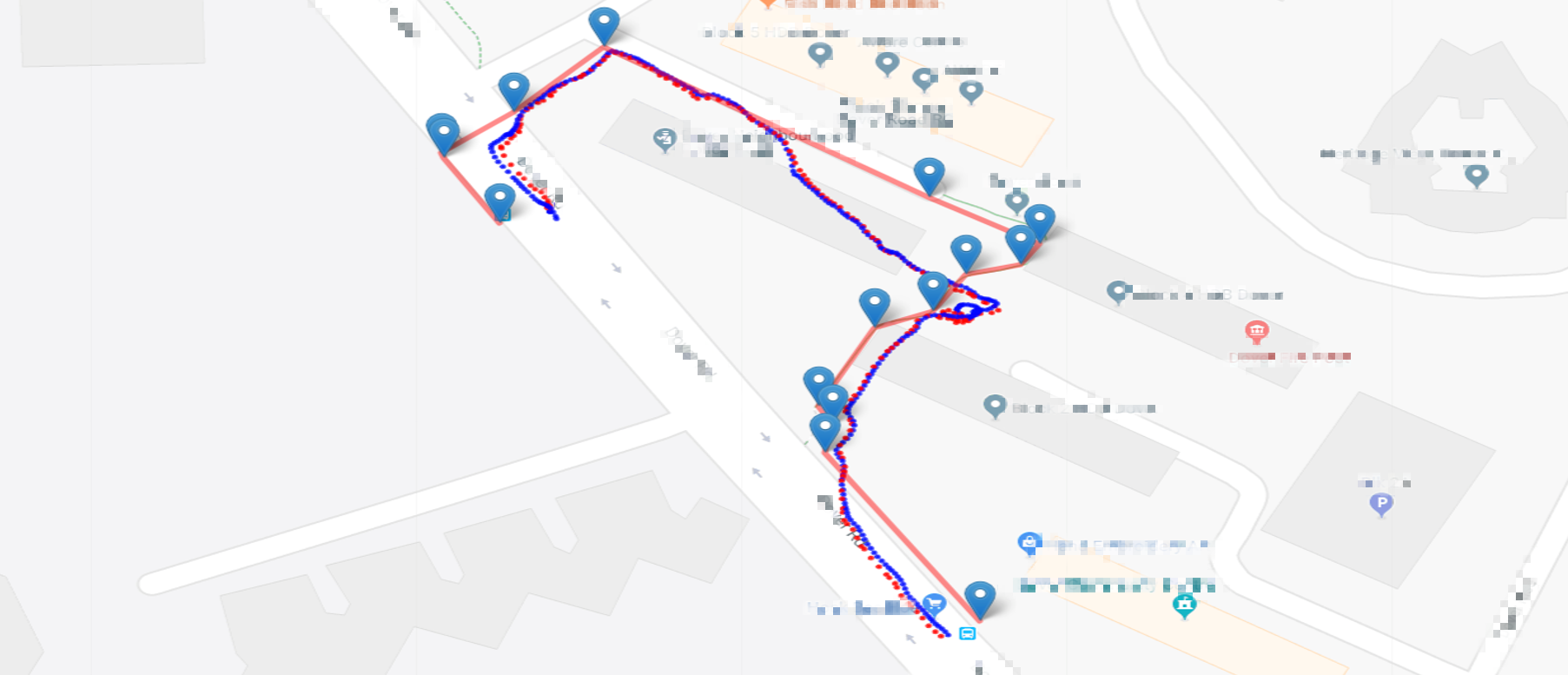}}
             }\\
      \subfigure[Error estimation for trace 1.]
      {\label{fig:trip_error_1}\includegraphics[width=.23\textwidth]{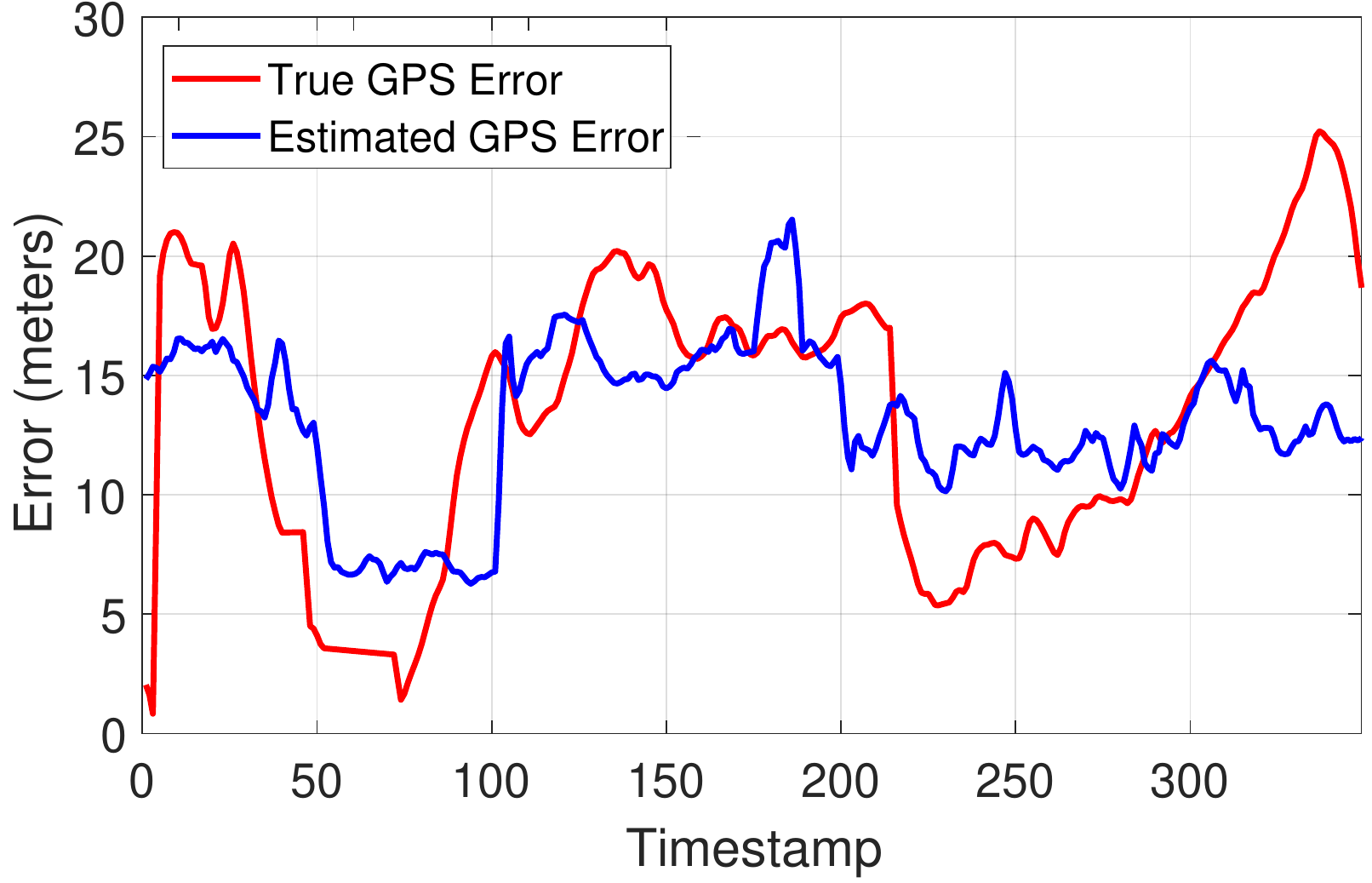}}
      \subfigure[Error estimation for trace 2.]{\label{fig:trip_error_2}\includegraphics[width=.23\textwidth]{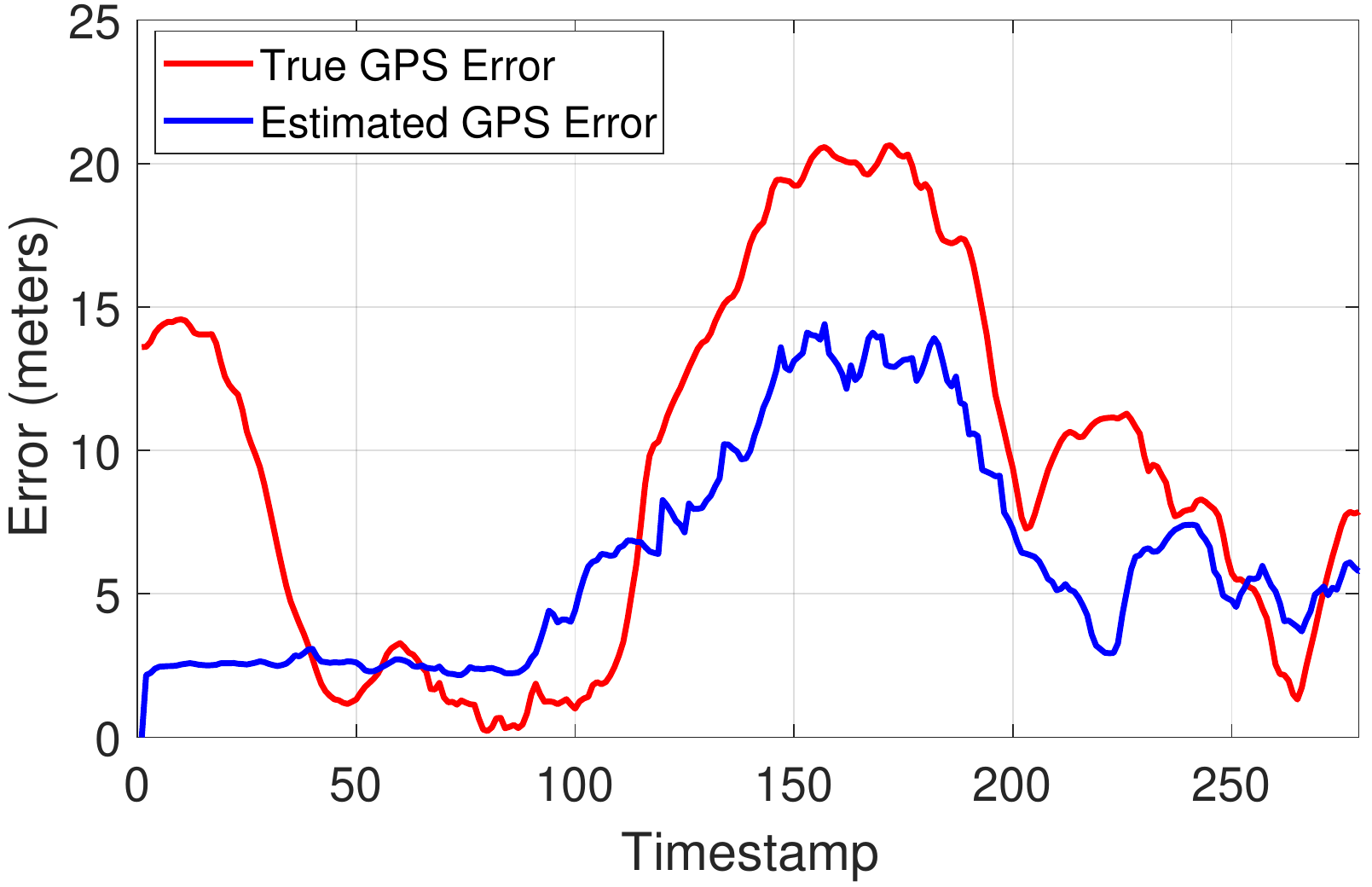}}
      \caption{Trace samples with error estimation.} 
      \label{fig:trace_error}
      \end{figure}

      We train the 3 models using representations generated from our antoencoder with the training traces, and test them by those generated with the testing traces. To compare with CRBM embedding discussed in Section~\ref{sssc:crbm}, we conduct the same procedure using the CRBM embedding model. The mean and median errors for all testing data are shown in \textbf{Figure~\ref{fig:error_comparison}}. DNN-R performs the best out of 3 models, achieving a median difference of 3.65 meters and a mean difference of 5.18 meters with the true errors. Our model generally performs better than CRBM except for RNN, but both models perform the worst in this case as they tend to over-fit the sequence pattern. To showcase the change of errors along a trace, 2 sample traces are also shown in \textbf{Figure~\ref{fig:trace_error}}, along with the comparison between the true error sequence and the estimated one based on our model using DNN-R. Both sequences generally capture the same error trends, though the magnitude may not perfectly match each other.

      To further evaluate the performance of our representation in estimating the GPS error, we compare with error estimators provided by Android system. Instead of an absolute estimated error range, both Android system~\cite{AndroidLocationApi} and Google~\cite{GoogleFusedApi} (yes, they are different APIs) provide an indicator called \textit{horizontal accuracy}, defined as the radius of 68\% confidence. In other words, there is a 68\% probability that the true location is inside the circle centered at reported location with a radius indicating the accuracy. To make our results comparable to it, we transform our absolute estimated error into a similar ``accuracy indicator'' by getting the 68\% percentile over the error distribution within past 10 seconds. 

\begin{figure}[t]
      \centering
        \subfigure[Mean probability of error falling within the confidence range at different time resolutions.] {\label{fig:mean_error}\includegraphics[width=.23\textwidth]{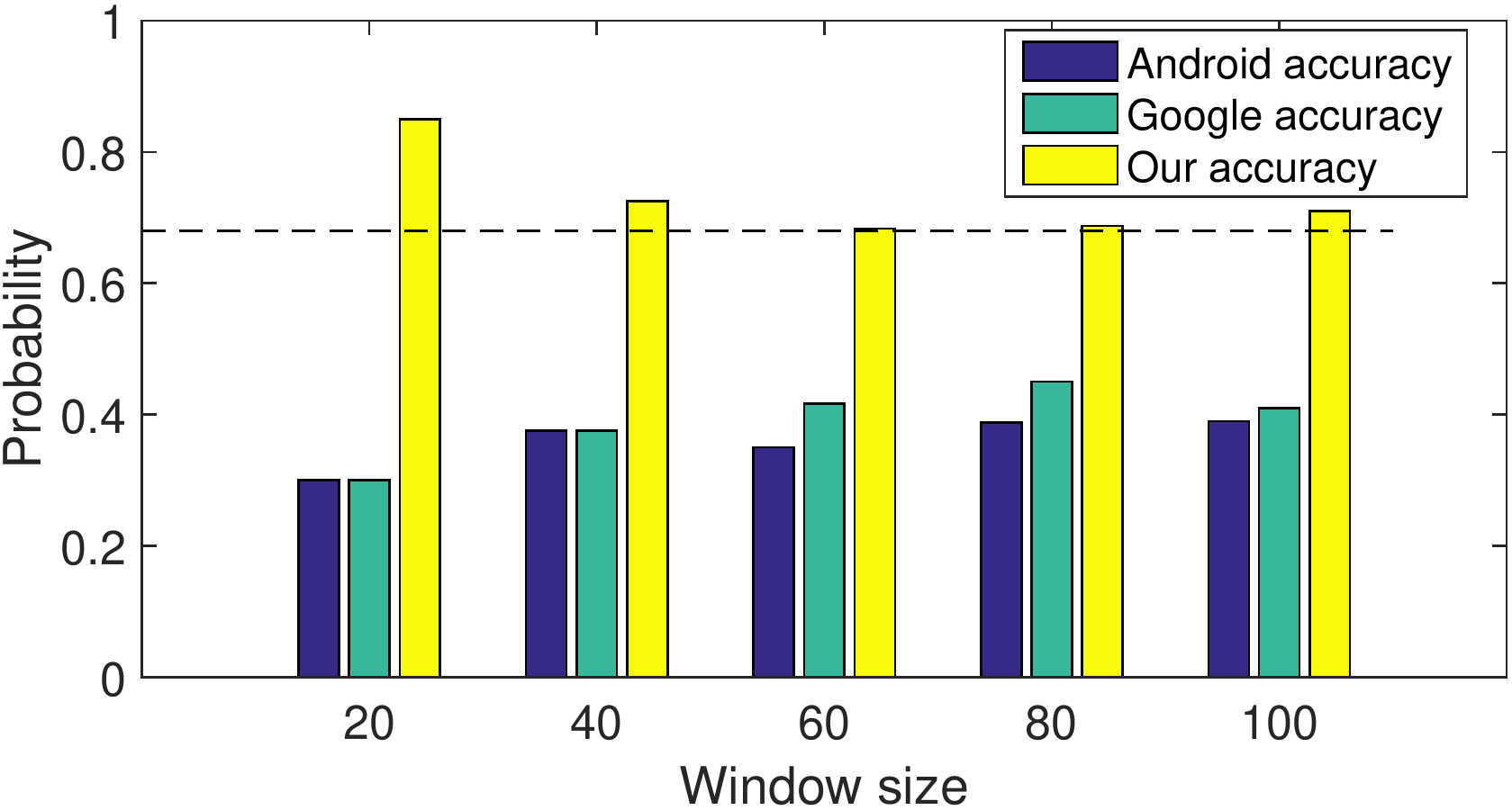}}
        \subfigure[Median probability of error falling within the confidence range at different time resolutions.]{\label{fig:median_error}\includegraphics[width=.23\textwidth]{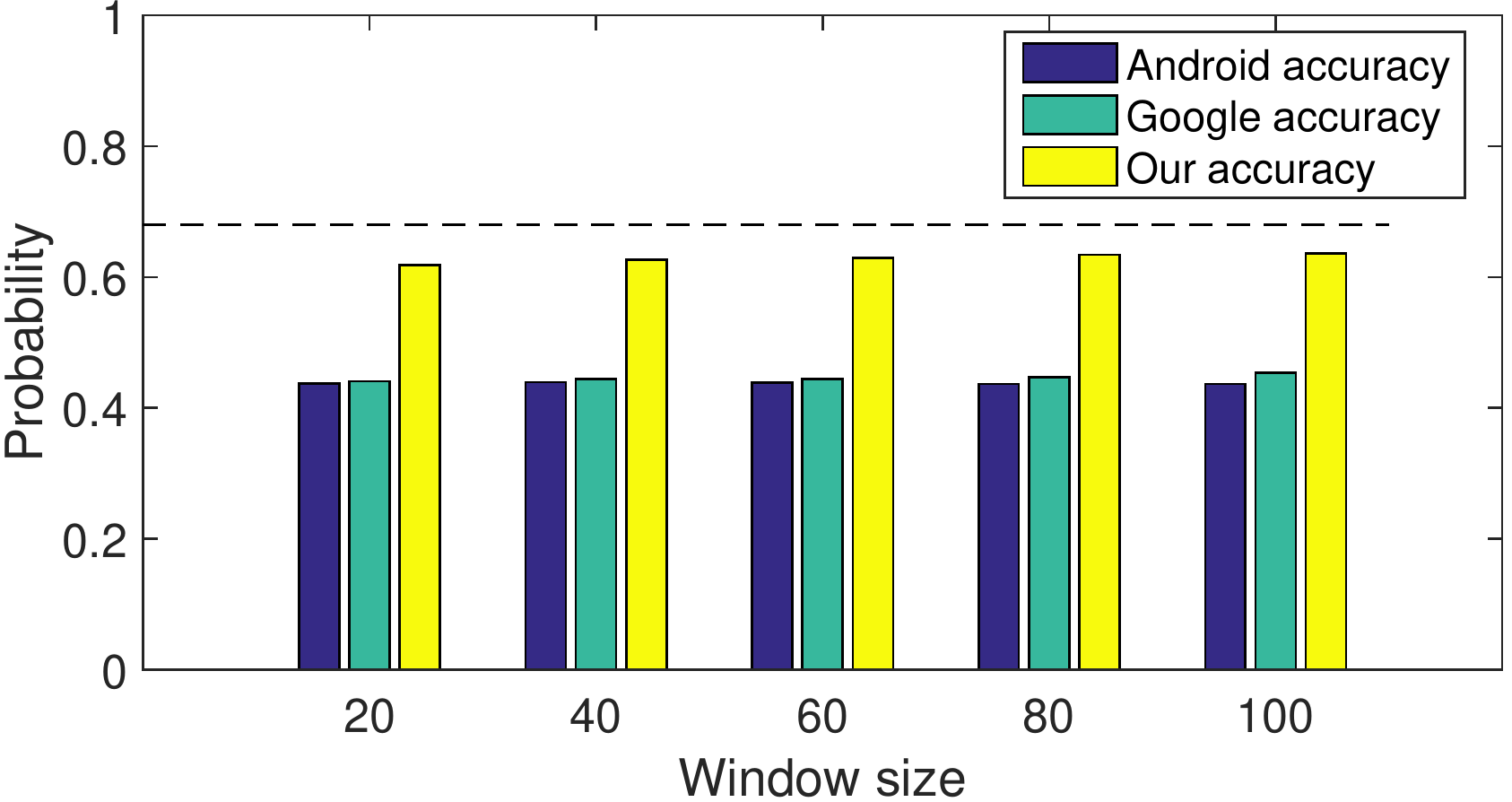}}
      \caption{Performance of localization accuracy indicator.}
      \label{fig:confidence_comparison}
      \end{figure}

      Based on the derived ``accuracy indicator'' for our model and the ``horizontal accuracy'' from Android system, we compute the probability that the true GPS error is smaller than each accuracy range within a sliding window. A probability close to 68\% implies an adequate accuracy indicator for localization performance. Larger than 68\% implies an over-conservative estimation, while smaller than 68\% implies an over-confident estimation. We also measure the probability with different sliding window sizes. \textbf{Figure~\ref{fig:confidence_comparison}} reports the mean and median probabilities over all testing trips for different sliding window sizes, and it makes a comparison among accuracy indicators derived from our model, Android~\cite{AndroidLocationApi}, and Google~\cite{GoogleFusedApi}.  All models give a relatively stable median probability and mean probability over different window sizes, indicating that the estimators perform in a consistent manner. However, the indicator delivered by our model stays almost around 68\%, while the other two indicators always give over-confident estimations. This suggests that our model yields a more reliable location accuracy indicator compared with the other two systems.

  \subsection{Location Contexts Semantic Analysis}
    To evaluate the effectiveness of representation on characterizing different contexts, we firstly build up a context database by adopting a bottom-up clustering process (i.e., lower-level clustering followed by upper-level clustering) with DBSCAN on all training trips. For 20 training trips, we finally get 86 lower-level clusters, 7 upper-level clusters and 14 small clusters remained as noise at upper level given their unique satellite patterns. To provide ground truth for training set, we manually assign predefined labels as described in Section \ref{sssc:data_collection} to all the representations in each clusters. For noise points detected through DBSCAN at lower-level clustering, we just label them as ``Noise''.
    \begin{figure}[t]
    \centering
    \subfigure[Success rate vs threshold.]{\label{fig:cluster_vs_threshold}\includegraphics[width=.23\textwidth]{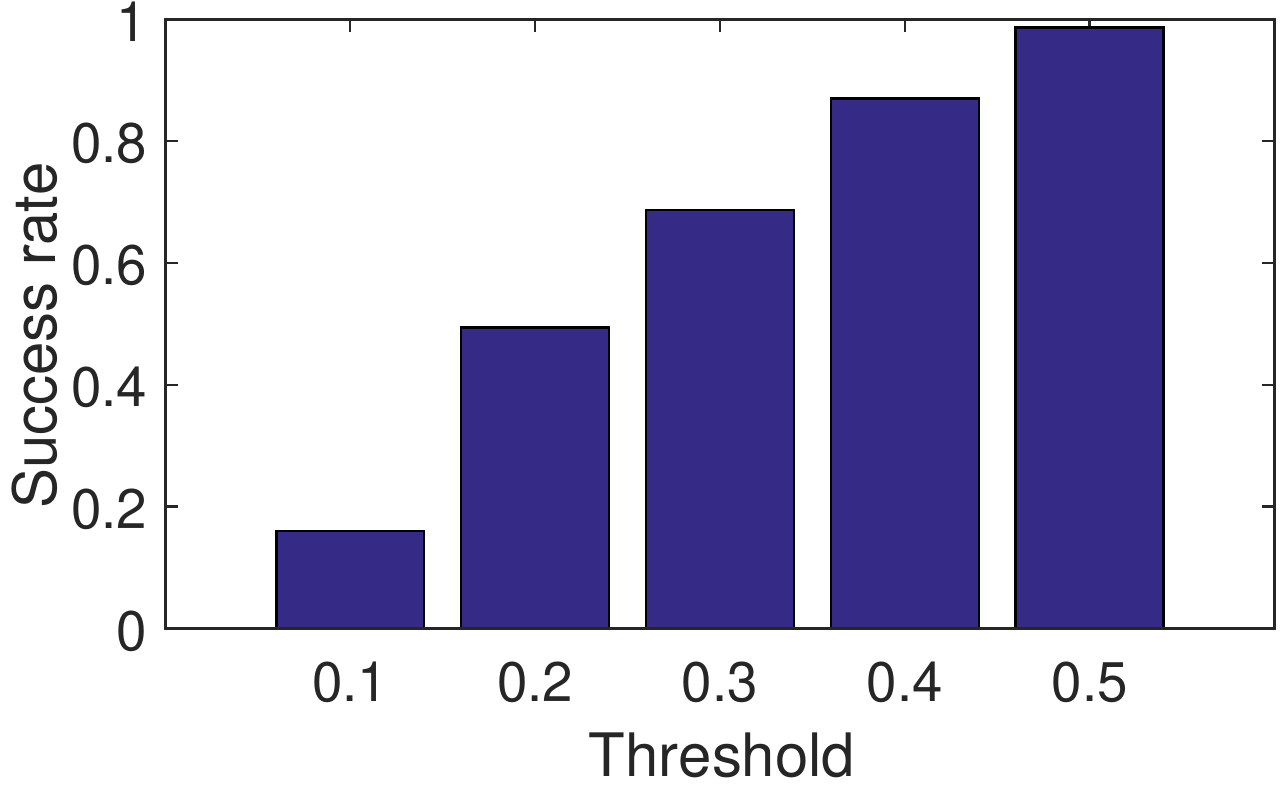}}
    \subfigure[Accuracy vs k for kNN search] {\label{fig:cluster_vs_k}\includegraphics[width=.238\textwidth]{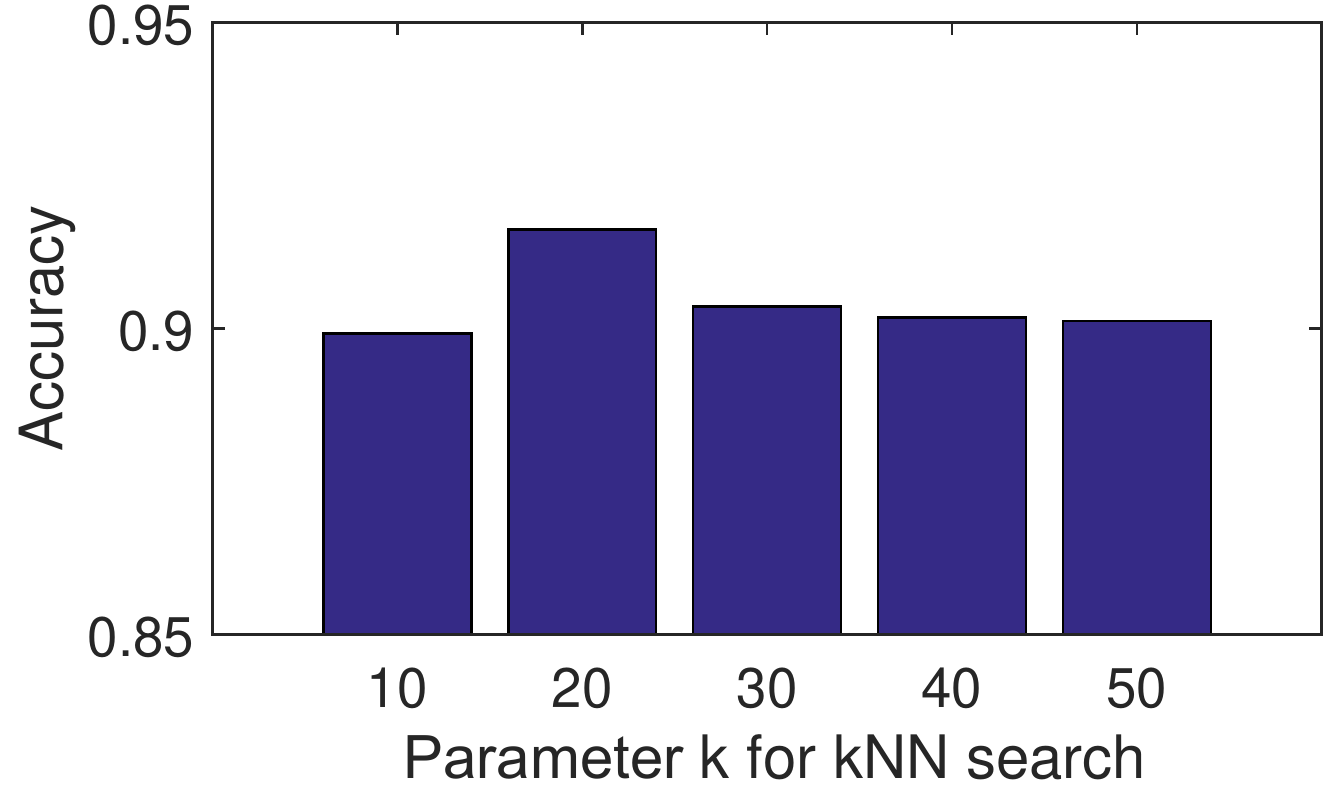}}
    \caption{Clustering performance with various threshold and k.} 
    \label{fig:cluster}
    \end{figure}

    At the query phase, we conduct the top-down search within our context database. We assign each representation with the same detailed context label of its $k$ nearest neighbors using cosine distance as metric. If the nearest distance is larger than our predefined threshold, we record it as a ``failure case''. We manually examine the assigned labels based on its corresponding location for validation. As shown in \textbf{Figure~\ref{fig:cluster}}, we firstly evaluate the success rate given different thresholds. The success rate of query increases when the threshold increases, and a threshold of 0.5 can cover 98.6\% of queries. We then evaluate classification accuracy under threshold of 0.5 for different $k$ values, and achieve the highest classification accuracy of 91.62\% when $k=20$; the accuracy at different $k$ values are all above 90\%. The results show that our model can successfully recognize most contexts given a sufficiently and well-labelled database.
    \begin{figure}[b]
    \centering 
    \parbox{\textwidth}{
          \parbox{.24\textwidth}{\center\includegraphics[width=.23\textwidth]{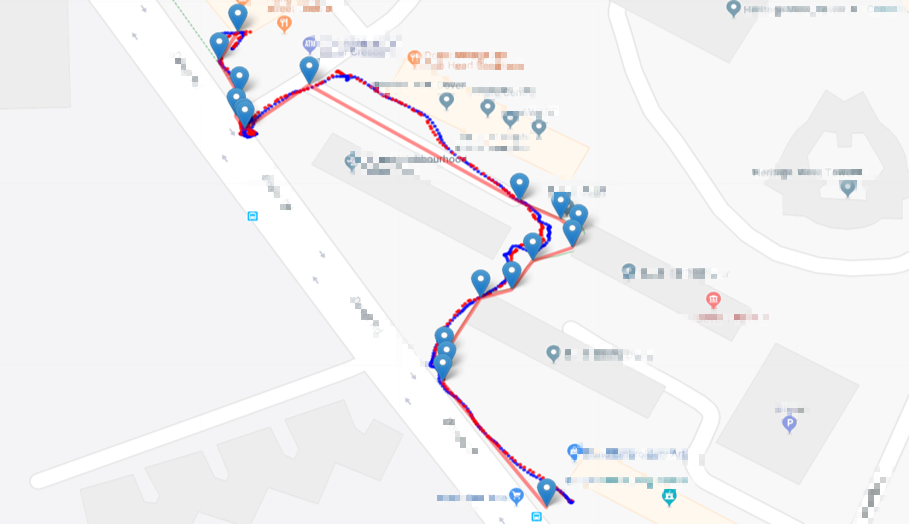}}
        \parbox{.24\textwidth}{\center\includegraphics[width=.23\textwidth]{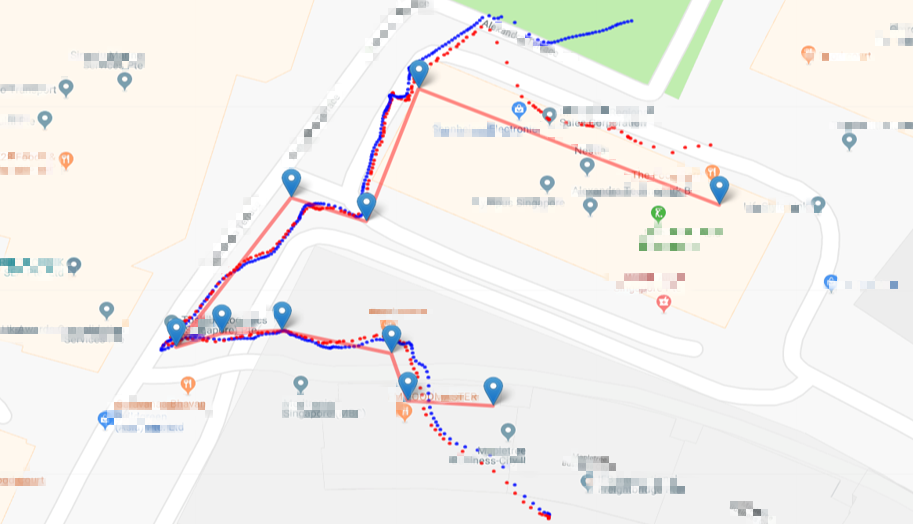}}
           }\\
    \subfigure[Label result for sample trace 1.]
    {\label{fig:trip_cluster_1}\includegraphics[width=.23\textwidth]{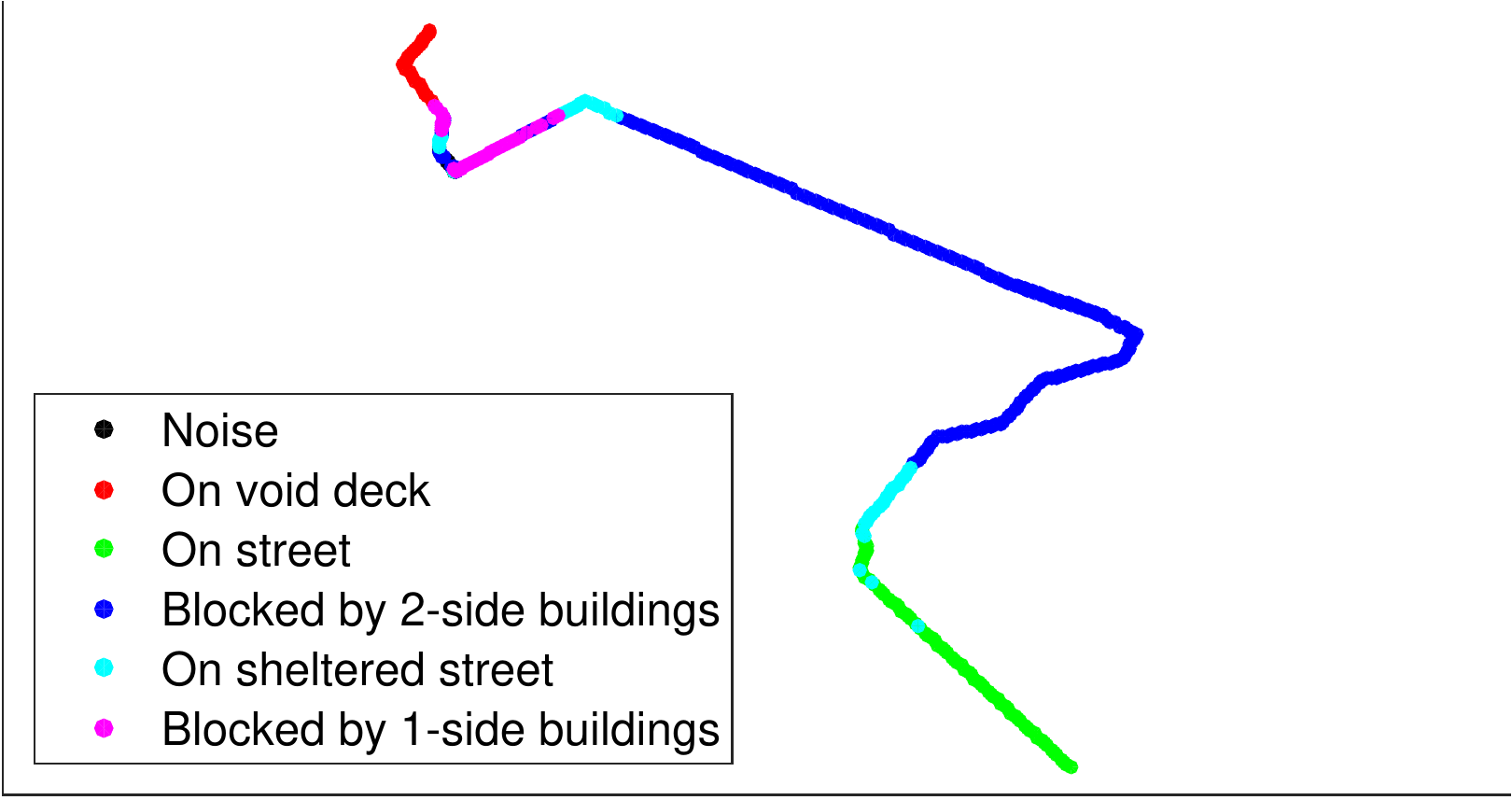}} \hspace{3pt}
    \subfigure[Label result for sample trace 2.]{\label{fig:trip_cluster_2}\includegraphics[width=.23\textwidth]{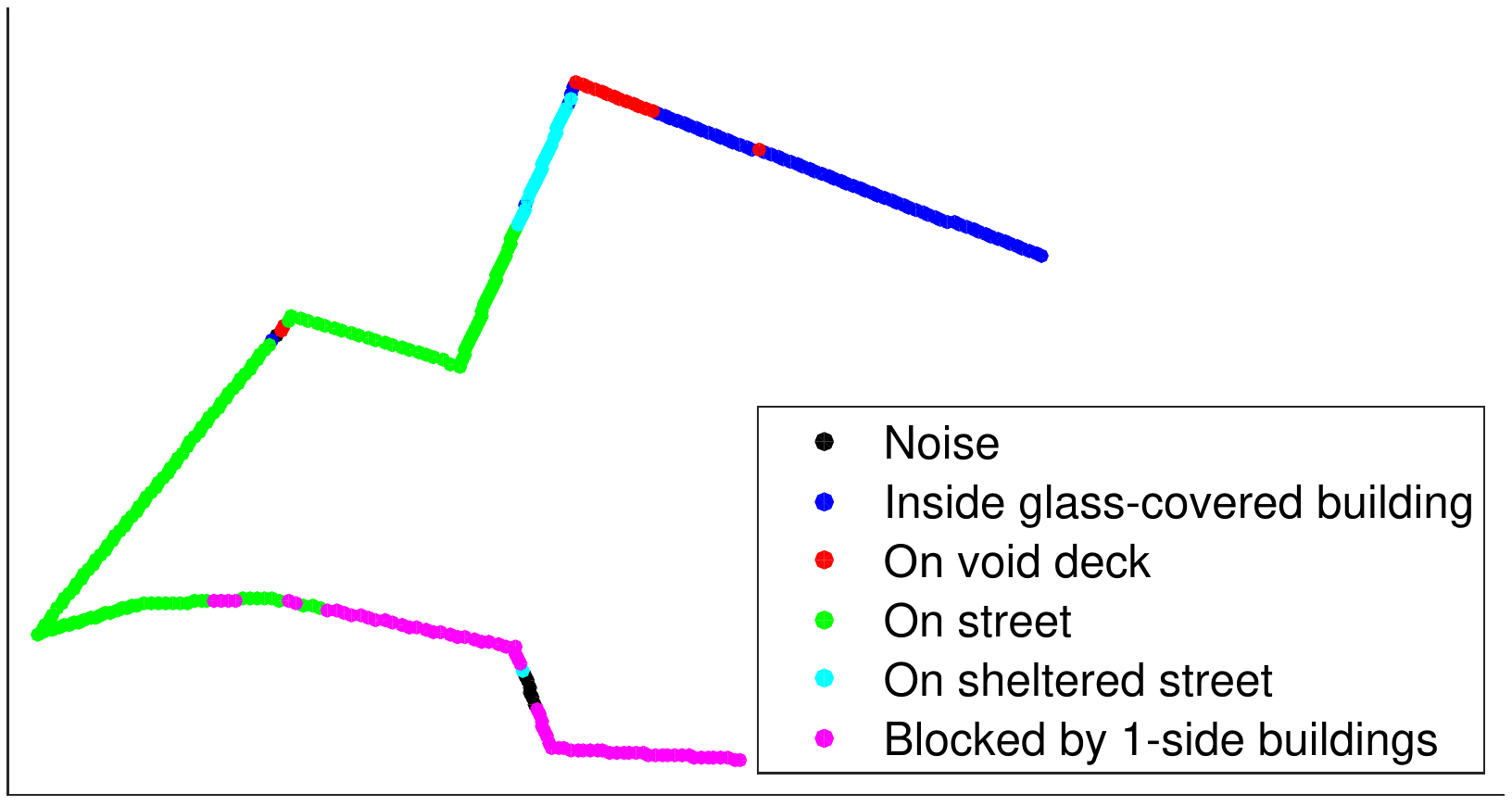}}
    \caption{Trace samples with context label results.}
    \label{fig:trace_label}
    \end{figure}

    To better showcase the performance in context semantic analysis, we demonstrate the result of two testing traces along with their snapshots taken from Google Maps in \textbf{Figure~\ref{fig:trace_label}}. Most of the location points are correctly labelled except some noise points and a mismatch between ``on void deck'' and ``on sheltered street'' in both traces given that two scenarios share quite similar contexts and hard to be distinguished.

  \subsection{Comparing with IO-Detectors}\label{ssec:energy}
    Directly comparing with earlier proposals on IO-detection~\cite{Zhou-SenSys12,Radu-SenSys14,Chen-INFOCOM2017} in terms of detection accuracy is impossible, as we offer a much finer granularity in context profiling than just a classification into three scenarios (i.e., indoor, outdoor, semi-outdoor), and our location context does not only include scenarios (see our definition in Section~\ref{ssec:case}). Earlier GPS-free proposals~\cite{Zhou-SenSys12,Radu-SenSys14} complain about the excessive delay incurred by a GPS-based approach, but the performance of commercial GPS module has been drastically improved since and we only retrieve raw GNSS measurements. Consequently, we have not experienced any excessive delay in detecting context changes during our experiments. As a result, the only meaningful comparison metric is energy efficiency. 

    To this end, we compare the energy consumption between our system and GPS-free IO-detection proposals~\cite{Zhou-SenSys12,Radu-SenSys14}, which rely mainly on magnetometer, cellular receiver, and light sensor. Therefore, we develop an Android application to collect magnetic field strength, cellular signal and light intensity at a frequency of 100Hz and install it on Huawei P10, in order to emulate the systems used in~\cite{Zhou-SenSys12,Radu-SenSys14}. By contrast, our application only collects raw GNSS measurements at a frequency of 1Hz. We measure the battery life when running each application with the screen set to the minimum brightness. As shown in \textbf{Table~\ref{tab:energy}}, our system consumes about 20\% more energy, largely comparable to the GPS-free solutions. Nonetheless, as our method achieves a much finer granularity in profiling location context compared with those earlier proposals, we believe that the slight disadvantage of our system in terms of energy consumption can be acceptable. In fact, there is still a room to further improve the energy efficiency of GNSS sensing, given that this is only a preliminary attempt to make use of this recently available sensing ability.
    \begin{table}[t]
    \centering
    \caption{Battery consumption comparison between two different sensor sets.}
      \begin{tabular}{|c|c|}
      \hline
      {\textbf{Sensor Set}} &{\textbf{Battery Life (Hours)}} \\ \hline
      Magnetometer + Cellular + Light &  29.2  \\ \hline
      Raw GNSS measurement listener&  24.2\\ \hline
      \end{tabular}
    \label{tab:energy}
    \end{table}
    \section{Conclusion} \label{sec:cc}
    In this paper, we have gone beyond well-studied indoor-outdoor detection and proposed to holistically profile the various location contexts in urban area. Based on extensive studies on raw GNSS measurements recently made available by Android 7, we have innovated in a new method to organize this unconventional data structure so that effective mining techniques can be applied. We have then engineered an autoencoder module to extract compact representations out of GNSS traces. In order to demonstrate the efficacy of this context profiling, we have showcased two applications: a localization error estimator better than that provided by Android/Google, and a context semantic database potential extensible to assist other applications, including the seamless integration of indoor and outdoor localization schemes.

    Our current work is still rather preliminary at this stage, mostly due to the fact that our GNSS data are only gathered by a few users. Therefore, this paper aims mainly to raise the attention of our community: should we gather such data in a more pervasive manner (e.g., crowdsensing incentivized by cloud offloading~\cite{Liu-SenSys2012}), we would be in a much better position to work towards full-fledged urban localization services.

\balance
\bibliographystyle{plain}
\bibliography{GNSS-LCP}

\end{document}